\documentstyle[aps,prd,preprint]{revtex}
\begin{document}
\draft
%%%%%%%%%%%%%%%%%%%%%%%%%%%%%%%%%%%%%%%%%%%%%%%%%%%%%%%%%%%%%%%%%%%%%%%%
\title{ van Vleck determinants:\\ traversable wormhole spacetimes}
\author{Matt Visser\cite{e-mail}}
\address{Physics Department, Washington University, St. Louis,
         Missouri 63130-4899}
\date{5 October 1993}
\maketitle
\begin{abstract}
Prompted by Hawking's {\em chronology protection conjecture}, and
by various other questions regarding the putative existence,
stability, and chronological properties of traversable wormholes,
a number of authors have presented calculations of the renormalized
stress--energy tensor in wormhole spacetimes.  In particular, the
use of point--splitting techniques leads to expressions that contain
the van Vleck determinant as a common prefactor.  Recent technical
advances permit one to undertake extensive computations of the van
Vleck determinant in traversable wormhole spacetimes --- at least
in the short--throat flat--space approximation.  This paper presents
several such computations for various model spacetimes. Implications
with regard to Hawking's chronology protection conjecture are
discussed.

In particular, any attempt to transform a single isolated wormhole
into a time machine results in large vacuum polarization effects.
These vacuum polarization effects are sufficient to disrupt the
internal  structure of the wormhole long before the onset of Planck
scale physics, and before the onset of time travel.  Thus for
isolated wormholes, vacuum polarization effects are sufficient to
enforce Hawking's chronology protection conjecture.

On the other hand, it is possible to conceive of a putative time
machine built out of two or more wormholes, each of which taken in
isolation is not itself a time machine. Such ``Roman configurations''
are much more subtle to analyse. For ``reasonable'' configurations
(traversable by humans) the vacuum polarization effects in such
multiple wormhole putative time machines become large long before
the onset of Planck scale physics. The disruption scale for would
be ``traversable time machines'' is well above the Planck scale.
On the other hand, for some particularly bizarre configurations
(not traversable by humans) the vacuum polarization effects can be
arranged to be arbitrarily small at the onset of Planck scale
physics.  This indicates that the disruption scale has been pushed
down into the Planck slop.  This is mildly disturbing.  Ultimately,
for these bizarre configurations, questions regarding the truth or
falsity of Hawking's chronology protection can only be addressed
by entering the uncharted wastelands of full fledged quantum gravity.
\end{abstract}

%%%%%%%%%%%%%%%%%%%%%%%%%%%%%%%%%%%%%%%%%%%%%%%%%%%%%%%%%%%%%%%%%%%%%%%%
\narrowtext
%%%%%%%%%%%%%%%%%%%%%%%%%%%%%%%%%%%%%%%%%%%%%%%%%%%%%%%%%%%%%%%%%%%%%%%%
\newpage
%%%%%%%%%%%%%%%%%%%%%%%%%%%%%%%%%%%%%%%%%%%%%%%%%%%%%%%%%%%%%%%%%%%%%%%%
\section{INTRODUCTION}
%%%%%%%%%%%%%%%%%%%%%%%%%%%%%%%%%%%%%%%%%%%%%%%%%%%%%%%%%%%%%%%%%%%%%%%%

Hawking's recently formulated chronology protection
conjecture~\cite{Hawking-I,Hawking-II} has prompted a considerable
amount of activity.  That conjecture, and various other issues
regarding the existence and putative stability of traversable
wormholes~\cite{Morris-Thorne,MTY,Visser89a,Visser89b}, have led
several authors to present calculations of the vacuum expectation
value of the quantum stress--energy tensor in various spacetimes
containing traversable wormholes
\cite{Kim-Thorne,Klinkhammer,Visser-CPC,Visser-Structure,GR13,Lyut}.
Generically one encounters expressions of the form
\begin{equation}
\langle 0|T_{\mu\nu}(x)|0\rangle
\approx \hbar \; \mathop{{\sum}'}_\gamma
  {\Delta_\gamma(x,x)^{1/2}\over\pi^2 s_\gamma(x,x)^4}
   \; t_{\mu\nu}(x;\gamma).
\label{stress-tensor}
\end{equation}
The sum runs over all nontrivial geodesics connecting the point
$x$ to itself. The symbol $s_\gamma(x,x)$ denotes the length of
the aforementioned geodesic. The tensor $t_{\mu\nu}(x;\gamma)$ is
a dimensionless object that is built up out of the spacetime metric
and the tangent vectors to $\gamma$~\cite{Visser-Structure}.
Furthermore $\Delta_\gamma(x,x)$ denotes the van Vleck determinant
\cite{vanVleck,Morette,Pauli,deWitt-I,deWitt-II,deWitt-III,deWitt-IV}.
The overall size of this prefactor is a key ingredient in governing
whether or not semiclassical quantum effects are sufficient to
enforce Hawking's chronology protection conjecture
\cite{Hawking-I,Hawking-II,Kim-Thorne,Visser-CPC,Visser-Structure,GR13,Lyut}.
To get a handle on what is going on, note that a good parameterization
of the magnitude of the quantum back reaction is obtained by
considering the scalar invariant
\begin{equation}
{\cal T} \equiv \sqrt{ \langle 0|T_{\mu\nu}(x)|0 \rangle \;
\langle 0|T^{\mu\nu}(x)|0 \rangle }.
\end{equation}
Then, if one of the closed self linking geodesics (say $\gamma'$)
dominates over the others, one may approximate
\begin{equation}
{\cal T} \approx \hbar \;
{\Delta_{\gamma'}(x,x)^{1/2}\over\pi^2 s_{\gamma'}(x,x)^4}.
\end{equation}
The essential point is that the semiclassical quantum back reaction
is directly proportional to the square root of the van Vleck
determinant. While the above formula is valid at any point $x$ in
the spacetime, it is most useful to consider a point $x$ on the
throat of the wormhole. The size of the back reaction should be
compared to the quantity of stress--energy required to hold the
wormhole throat open in the first place. At the throat of a
wormhole whose mouth is of radius $R$ the Einstein equations imply
\begin{equation}
{\cal T}_0 \equiv \sqrt{ T_{\mu\nu}\; T^{\mu\nu} }
\approx {\hbar\over\ell_P^2 R^2}.
\end{equation}
Once ${\cal T} \gg {\cal T}_0$ one has what I feel is clear and
convincing evidence that the semiclassical vacuum polarization
disrupts the internal structure of the wormhole~\cite{Visser-Structure}.
(Other authors may choose to disagree with me on this issue). This
disruption occurs at
\begin{equation}
s\vert_{\rm disrupt}
\approx \Delta_{\gamma'}^{1/8} \sqrt{\pi \ell_P R}.
\end{equation}
Since $s_\gamma(x,x) \to 0$ is the signal for the onset of time
travel, it will {\em always} be the case that this occurs before
the time machine has a chance to form.  The real issue is whether
the overwhelming of the wormhole's internal structure occurs before
or after one reaches the Planck regime.

If ${\cal T} \gg {\cal T}_0$ while $s_\gamma(x,x) \gg \ell_P$, then
semiclassical vacuum polarization effects overwhelm the wormhole's
internal structure in a parameter regime where one still expects
the semiclassical approximation to hold. One may thus safely assert
that at least in this parameter regime, semiclassical vacuum
polarization effects enforce Hawking's chronology protection
conjecture.

If ${\cal T} \gg {\cal T}_0$ does not occur until $s_\gamma(x,x) <
\ell_P$, then one cannot conclude that  time machine formation
succeeds. All that one can safely conclude is that the uncharted
wastelands of full fledged quantum gravity have been entered.

A central issue in investigating these issues is thus the evaluation
of $\Delta_{\gamma}(x,x)$ for closed geodesics $\gamma$ with base
point $x$ located on the throat of the wormhole.  In fact, I shall
try to be more general than that.

Recent technical advances~\cite{Visser-vanVleck} permit radical
improvements in the computation of the van Vleck determinant.
Basic tools employed are the reformulation of the van Vleck
determinant in terms of tidal focussing effects; and consequent
formal expansion for the van Vleck determinant in terms of the
Riemann tensor.  If the Riemann curvature is confined to relatively
thin layers a suitable extension of the usual thin--shell formalism
permits reduction of this formal expansion to a finite number of
terms.  The number $N$ of terms occurring in this expansion is just
the number of times the closed geodesic $\gamma$ intersects a layer
of high curvature.  Several variations on this theme are presented.

\underbar{Notation:} Adopt units where $c\equiv 1$, but all other
quantities retain their usual dimensionalities. In particular
$G=\hbar/m_P^2 = \ell_P^2/\hbar$. The metric signature is taken to
be $(-,+,+,+)$. General conventions follow Misner, Thorne,
and Wheeler~\cite{MTW}.

%%%%%%%%%%%%%%%%%%%%%%%%%%%%%%%%%%%%%%%%%%%%%%%%%%%%%%%%%%%%%%%%%%%%%%%%%%%%%%
\section{THE VAN VLECK DETERMINANT}
%%%%%%%%%%%%%%%%%%%%%%%%%%%%%%%%%%%%%%%%%%%%%%%%%%%%%%%%%%%%%%%%%%%%%%%%%%%%%%

The primary definition of the (scalarized) van Vleck determinant is
in terms of the second derivatives of the arc--length function
\begin{equation}
\Delta_\gamma(x,y) \equiv
\pm {1\over2} {1\over\sqrt{g(x)g(y)}}
\det\left\{
{\partial^2 [s_\gamma(x,y)]^2
\over\partial x^\mu \, \partial y^\nu}
\right\}_.
\end{equation}
The symbol $s_\gamma(x,y)$ denotes the arc--length along the geodesic
$\gamma$.

In a Lorentzian spacetime (or a Riemannian manifold) the van Vleck
determinant can be reformulated in terms of tidal focussing effects
\cite{Visser-vanVleck}. Define
\begin{equation}
\Delta_\gamma(x,y) = \det\left[ s_\gamma(x,y) A^{-1} \right].
\end{equation}
Here the matrix $A$ describes the evolution of a set of Jacobi
fields along the geodesic $\gamma$. In terms of components defined
by Fermi-Walker transporting a set of basis vectors along the
geodesic $\gamma$, the matrix $A$ is governed by the second-order
differential equation~\cite{Visser-vanVleck,Hawking-Ellis}
\begin{equation}
{d^2\over ds^2} A^\mu{}_\nu (s)
= - \left( R^\mu{}_{\alpha\sigma\beta} \; t^\alpha t^\beta \right)
A^\sigma{}_\nu.
\label{tidal}
\end{equation}
The relevant boundary conditions are that $A^\mu{}_\nu(s) \to s\,
\delta^\mu{}_\nu$ as $s\to0$. Introducing the one--dimensional Green
function $G_R(s_f,s_i) = \{s_f-s_i\} \Theta(s_f-s_i)$, this may be
reformulated as an integral equation
\begin{equation}
A^\mu{}_\nu(s) =
s\, \delta^\mu{}_\nu  - \int_0^s   G_R(s,s')
  \left( R^\mu{}_{\alpha\sigma\beta} t^\alpha t^\beta \right)
   A^\sigma{}_\nu(s') ds'.
\label{integral}
\end{equation}
This integral equation may be solved by iteration. Defining
$Q^\mu{}_\nu \equiv - \left( R^\mu{}_{\alpha\nu\beta} t^\alpha t^\beta
\right)$, and suppressing the explicit integration by regarding
$G_R(s,s')$, multiplication by $Q(s)$, and multiplication by $s$,
as functional operators
\begin{equation}
 A =  \left(I + [G_R Q] + [G_R Q]^2 + \cdots \right) \{sI\}.
\label{formal}
\end{equation}
The formalism as presented here applies equally well to spacelike
or timelike geodesics. Lightlike geodesics require a little extra
technical fiddling --- for details see reference~\cite{Visser-vanVleck}.

%%%%%%%%%%%%%%%%%%%%%%%%%%%%%%%%%%%%%%%%%%%%%%%%%%%%%%%%%%%%%%%%%%%%%%%%%%%%%%
\section{WORMHOLE GEOMETRY: FIRST APPROXIMATION}
%%%%%%%%%%%%%%%%%%%%%%%%%%%%%%%%%%%%%%%%%%%%%%%%%%%%%%%%%%%%%%%%%%%%%%%%%%%%%%

Consider a traversable wormhole: Assume that the throat of
the wormhole is very short, and that curvature in the region outside
the mouth of the wormhole is relatively weak. Such a wormhole can
be idealised by considering Minkowski space with two regions excised,
and then identifying the boundaries of those regions in some suitable
manner. The Riemann tensor for such an idealised geometry is
identically zero everywhere except at the wormhole mouths/throat
where the identification procedure takes place~\cite{Visser89a,Visser89b}.
Generically, there will be an infinitesimally thin layer of exotic
matter present at the mouth of the wormhole~\cite{Morris-Thorne,MTY}.

%%%%%%%%%%%%%%%%%%%%%%%%%%%%%%%%%%%%%%%%%%%%%%%%%%%%%%%%%%%%%%%%%%%%%%%%%%%%%%
\subsection{First approximation: first attempt}
%%%%%%%%%%%%%%%%%%%%%%%%%%%%%%%%%%%%%%%%%%%%%%%%%%%%%%%%%%%%%%%%%%%%%%%%%%%%%%

Consider now a geodesic that wraps through the mouth of the wormhole
a total of $N$ times. In this short--throat flat--space approximation
one has
\begin{equation}
Q^\mu{}_\nu(s)
\equiv - \left( R^\mu{}_{\alpha\nu\beta} t^\alpha t^\beta \right)
=  \sum_{n=1}^{N} \delta(s-s_n) [q(n)^\mu{}_\nu].
\end{equation}
This greatly simplifies the various terms in the expansion for the
van Vleck determinant. In particular
\begin{equation}
[G_R Q]\{sI\}
= \sum_{n=1}^{N} (s-s_n) [q(n)^\mu{}_\nu] s_n.
\end{equation}
Furthermore
\begin{equation}
{[G_R Q]}^2\{sI\}
= \sum_{n=2}^{N} \sum_{m=1}^{n-1} (s-s_n)
[q(n)^\mu{}_\rho] (s_n-s_m) [q(m)^\mu{}_\nu] s_m.
\end{equation}
While, using the symbol $\bullet$ to denote a generic dummy index,
\begin{eqnarray}
[G_R Q]^{N-1}\{sI\} =
\sum_{n=1}^{N}
&& ( s-s_N ) [q(N)^\mu{}_{\bullet}] (s_N-s_{N-1})  \cdots \nonumber\\
&& \qquad  [q(n+1)^\bullet{}_\bullet] (s_{n+1} - s_{n-1})
           [q(n-1)^\bullet{}_\bullet]
\cdots  \nonumber\\
&& \qquad \qquad ( s_2-s_1 )[q(1)^{\bullet}{}_\nu] s_1.
\end{eqnarray}
Ultimately
\begin{equation}
[G_R Q]^N\{sI\} =
(s-s_N) [q(N)^\mu{}_{\bullet}] (s_N-s_{N-1})
\cdots  (s_2-s_1) [q(1)^{\bullet}{}_\nu] s_1.
\end{equation}
Due to the presence of a sufficient number of Heavyside functions,
higher powers of $[G_R Q]$ vanish: $[G_R Q]^{N+n}\{sI\}
\Rightarrow 0$ for $n>0$.   Note that the formal expansion
(\ref{formal}) now terminates in a finite number of steps
\begin{equation}
 A =  \left(I + [G_R Q] + [G_R Q]^2 + \cdots + [G_R Q]^N \right) \{sI\}.
\end{equation}
Here $N$ denotes the total number of trips through the wormhole.
Observe, either from the above, or from the differential equation
(\ref{tidal}), that in this type of geometry the matrix $A(s)$ is
a piecewise linear continuous matrix function of arc length.
Consequently the reciprocal of the van Vleck determinant
$\Delta_\gamma(s)^{-1} = \det\{A/s\}$ is piecewise a Laurent
polynomial in arc length~\cite{Visser-vanVleck}.

At this stage, I am deliberately leaving the matrices $[q(n)^\mu{}_\nu]$
as general as possible so as to obtain results that are to a large
extent independent of the particular details of the shape of the
wormhole mouths and/or the identification procedure adopted at the
wormhole mouths. Suitable specializations will be introduced in
due course.

%%%%%%%%%%%%%%%%%%%%%%%%%%%%%%%%%%%%%%%%%%%%%%%%%%%%%%%%%%%%%%%%%%%%%%%%%%%%%%
\subsection{First approximation: single pass}
%%%%%%%%%%%%%%%%%%%%%%%%%%%%%%%%%%%%%%%%%%%%%%%%%%%%%%%%%%%%%%%%%%%%%%%%%%%%%%

For a geodesic that makes only a single pass through the mouth of
the wormhole ($N=1$), one easily derives an exact closed form expression
\begin{equation}
A^\mu{}_\nu(s) =
s \delta^\mu{}_\nu + \Theta(s-s_1) \{ (s-s_1) [q(1)^\mu{}_\nu] s_1 \}.
\end{equation}
This result may be obtained either from the formal manipulations of
the preceding section, or from direct integration of the tidal
equation. This leads to a closed form for the van
Vleck determinant. For $s\geq s_1$:
\begin{equation}
\Delta_\gamma(s)^{-1}
= \det\left( \delta^\mu{}_\nu
            + {(s-s_1)s_1\over s} [q(1)^\mu{}_\nu] \right)_.
\end{equation}
Particularize to a closed geodesic, so that $x=y$. Define $s_+
= s_1$, $s_- = s - s_1$, and let $s=s_+ + s_-$ denote the total
arc length.  For convenience, one may set $[q^\mu{}_\nu] \equiv
[q(1)^\mu{}_\nu]$. For such a closed geodesic
\begin{equation}
\Delta_\gamma(x,x)
= \det\left( \delta^\mu{}_\nu
            + {s_+ s_-\over s} [q^\mu{}_\nu] \right)^{-1}_.
\end{equation}
Suppose that the point of interest, $x=y$, lies near the throat of
the wormhole.  Then either $s_+\approx 0$ or $s_-\approx 0$.

\underbar{Lemma:} For any closed geodesic $\gamma$ that threads
the throat of the wormhole once, for any point $x$ near the throat
of the wormhole, the short--throat flat--space approximation yields:
\begin{equation}
\Delta_\gamma(x,x) = 1 - {s_+ s_-\over s} \hbox{\rm tr}[q]
                   + O[s_+^2 s_-^2 /s^2].
\end{equation}

At the throat itself , of course, this implies $\Delta_\gamma(x,x)
= 1$. This observation is already of some interest.
See~\cite{Visser-Structure} and the discussion section of this
paper.  This result is compatible with, and a generalization of,
the $N=1$ case of equation (48) of Kim and Thorne~\cite{Kim-Thorne}.

On the other hand, suppose that the point $x=y$ is far away from
the wormhole throat. The behaviour of the van Vleck determinant is
now governed by whether or not  the eigenvalues of $\{s_+ s_-/s\}
[q^\mu{}_\nu]$ ever become large compared to $1$. This depends on
the details of the wormhole's construction.

\underbar{Lemma:} For any closed geodesic $\gamma$ that threads
the throat of the wormhole once, for any point $x$ far away from
throat of the wormhole, the short--throat flat--space approximation
yields:
\begin{equation}
\Delta_\gamma(x,x) =
\left[{s\over s_+ s_-}\right]^{\#}  {\det}'[q]^{-1}
+ O\left[\left({s\over s_+ s_-}\right)^{\#-1}\right]_.
\end{equation}
Here $\#$ denotes the number of large eigenvalues of the matrix $\{s_+
s_-/s\}[q^\mu{}_\nu]$. Furthermore $\det'$ denotes the determinant
with small eigenvalues omitted.  If all of the eigenvalues are
small, then $\#=0$, $\det'[q]=1$, and consequently $\Delta_\gamma(x,x)
\approx 1$.

This is compatible with the $N=1$ case of equation (64) of Kim and
Thorne~\cite{Kim-Thorne}. The present result is derived in a much
more general context than their result but does not (yet) contain
as much information.

%%%%%%%%%%%%%%%%%%%%%%%%%%%%%%%%%%%%%%%%%%%%%%%%%%%%%%%%%%%%%%%%%%%%%%%%%%%%%%
\subsection{First approximation: double pass}
%%%%%%%%%%%%%%%%%%%%%%%%%%%%%%%%%%%%%%%%%%%%%%%%%%%%%%%%%%%%%%%%%%%%%%%%%%%%%%

For a geodesic that makes two passes through the mouth of the
wormhole ($N=2$), one has the closed form expression
\begin{eqnarray}
A^\mu{}_\nu(s)
&=&  s \delta^\mu{}_\nu  \nonumber\\
&& + \Theta(s-s_1)
   \left\{(s-s_1) [q(1)^\mu{}_\nu] s_1 \right\}
\nonumber\\
&& + \Theta(s-s_2)
    \left\{(s-s_2) [q(2)^\mu{}_\nu] s_2 \right\}
\nonumber\\
&&+ \Theta(s-s_2)
    \left\{(s-s_2) [q(2)^\mu{}_\rho] (s_2-s_1)
                             [q(1)^\rho{}_\nu] s_1 \right\}_.
\end{eqnarray}
As was the case previously, this result may be derived either from
the formal expansion, or from direct integration of the tidal
equation.

Now particularize to a closed geodesic, so that $x=y$. Define $s_+
= s_1$, $s_- = s - s_2$, and $s_0=s_2-s_1$. Then $s=s_+ + s_0 + s_-$
denotes the total arc length.  For convenience, one may set $[q_+]
= [q(1)]$, and $[q_-] = [q(2)]$. Thus, for a closed two--pass geodesic
\begin{equation}
\Delta_\gamma(x,x)^{-1}
= \det\left(
\delta^\mu{}_\nu
+  {s_+(s_- + s_0)\over s} [q_+^\mu{}_\nu]
+  {s_-(s_+ + s_0)\over s} [q_-^\mu{}_\nu]
+  {s_+ s_- s_0\over s}[q_-^\mu{}_\rho][q_+^\rho{}_\nu] \right)_.
\end{equation}

Now suppose the base point $x$ of the geodesic is near the throat of
the wormhole. For definiteness take $s_+\equiv s_1\approx0$, though
one could just as easily take $s_- \equiv s- s_2\approx 0$. Introduce
some new notation: $\epsilon \equiv s_1 \approx 0$, $s_{1\to2}\equiv
s_2 - s_1$, $s_{2\to 1} \equiv s_1 + s - s_2 \approx s- s_2$.

\underbar{Lemma:} For any closed geodesic $\gamma$ that threads
the throat of the wormhole twice, for any point $x$ near the throat
of the wormhole, the short--throat flat--space approximation yields:
\begin{equation}
\Delta_\gamma(x,x)^{-1}
= \det\left(
\delta^\mu{}_\nu
+  {s_{1\to2} \; s_{2\to1} \over s_{1\to2} + s_{2\to1} }[q(2)^\mu{}_\nu]
+  O[\epsilon] \right)_.
\end{equation}
Taking $\#$ to denote the number of large eigenvalues of the matrix
${s_{1\to2} \; s_{2\to1} \over s_{1\to2} + s_{2\to1} }[q(2)^\mu{}_\nu]$
one may approximate
\begin{equation}
\Delta_\gamma(x,x)^{-1}
\approx \left[
        {s_{1\to2} \; s_{2\to1} \over s_{1\to2} + s_{2\to1} }
        \right]^{\#}
{\det}' [q(2)^\mu{}_\nu].
\end{equation}

Far away from the wormhole throat the analysis is considerably more
delicate. The critical question in this case is whether or not any
of the eigenvalues of the matrix $\{(s_+ s_- s_0)/s\}[q_- q_+]$
are large.

\underbar{Lemma:} For any closed geodesic $\gamma$ that threads
the throat of the wormhole twice, if any of these eigenvalues are
large, the short--throat flat--space approximation yields:
\begin{equation}
\Delta_\gamma(x,x) =
\left[{s\over s_+ s_- s_0}\right]^{\#}  {\det}'[q_+q_-]^{-1}
+ O\left[\left({s\over s_+ s_- s_0}\right)^{\#-1}\right]_.
\end{equation}
As before, let $\#$ denote the number of large eigenvalues, this time
of the matrix $\{(s_+ s_- s_0)/s\}[q_+ q_-]$.   If all of
the eigenvalues are small, then one must revert to
\begin{equation}
\Delta_\gamma(x,x)^{-1}
\approx \det\left(
\delta^\mu{}_\nu
+  {s_+(s_- + s_0)\over s} [q_+^\mu{}_\nu]
+  {s_-(s_+ + s_0)\over s} [q_-^\mu{}_\nu]
\right)_.
\end{equation}

For the case of a double pass through the wormhole, this is about
as far as this general type of analysis can profitably be carried.
To improve and extend the calculations one needs more detailed
information about the matrices $[q(n)]$.

%%%%%%%%%%%%%%%%%%%%%%%%%%%%%%%%%%%%%%%%%%%%%%%%%%%%%%%%%%%%%%%%%%%%%%%%%%%%%%
\subsection{First approximation: multiple passes}
%%%%%%%%%%%%%%%%%%%%%%%%%%%%%%%%%%%%%%%%%%%%%%%%%%%%%%%%%%%%%%%%%%%%%%%%%%%%%%

If one were to retain a completely general geodesic, the algebraic
complexity involved in evaluating the van Vleck determinant would
quickly rise from the merely cumbersome to the absolutely prohibitive.
Accordingly, for this section only, I shall specialize to the case
of a completely smooth closed geodesic, that wraps around the
wormhole $N$ times.

Because I have assumed complete smoothness,
including smoothness at the base point $x$, the geodesic is a
$N$--fold overlay of a smooth once around the wormhole geodesic.
Thus the points where the geodesic intersects the wormhole mouth
are all identical and one has
\begin{equation}
[q^\mu{}_\nu] \equiv [q(1)^\mu{}_\nu] = [q(2)^\mu{}_\nu] = \cdots
 = [q(N)^\mu{}_\nu].
\end{equation}
Furthermore
\begin{equation}
s_0 \equiv s_N - s_{N-1} = \cdots = s_3 - s_2 = s_2 - s_1.
\end{equation}
Finally, define $s_+ = s_1$, and $s_- = s - s_N$. Since the total
length of the geodesic is $s = N s_0$ it follows that $s_+ + s_-
= s_0$. Indeed $s_n = s_+ + (n-1)s_0 = n s_0 - s_-$. Evaluating
the various terms in the expansion for the van Vleck determinant
will require a brief agony of combinatorics.

%%%%%%%%%%%%%%%%%%%%%%%%%%%%%%%%%%%%%%%%%%%%%%%%%%%%%%%%%%%%%%%%%%%%%%%%%%%%%%
\subsubsection{Zero--order term}
%%%%%%%%%%%%%%%%%%%%%%%%%%%%%%%%%%%%%%%%%%%%%%%%%%%%%%%%%%%%%%%%%%%%%%%%%%%%%%

The zeroth order term is just
\begin{equation}
[G_R Q]^0\{sI\} = s = N s_0.
\end{equation}

%%%%%%%%%%%%%%%%%%%%%%%%%%%%%%%%%%%%%%%%%%%%%%%%%%%%%%%%%%%%%%%%%%%%%%%%%%%%%%
\subsubsection{First--order term}
%%%%%%%%%%%%%%%%%%%%%%%%%%%%%%%%%%%%%%%%%%%%%%%%%%%%%%%%%%%%%%%%%%%%%%%%%%%%%%

It is a simple matter to see that
\begin{equation}
[G_R Q]\{sI\}
= \sum_{n=1}^{N} (s_- + [N-n]s_0)(s_+ + [n-1]s_0) \times
[q^\bullet{}_\bullet].
\end{equation}
Define $p_\pm= s_\pm/s_0$, so that $p_+ + p_-=1$. Extracting all
dimensionfull factors yields
\begin{equation}
[G_R Q]\{sI\}
= s_0 \; S(N,1) \; (s_0 [q^\bullet{}_\bullet]) .
\end{equation}
The sum can be rearranged to be
\begin{eqnarray}
S(N,1)
&=& \sum_{n=1}^{N}
\left( p_+ p_- +  \{p_+ [N-n] + p_- [n-1]\} + [N-n][n-1] \right)
\nonumber\\
&=& \sum_{n=1}^{N}
\left( p_+ p_- + [N-n] + [N-n][n-1] \right)
\nonumber\\
&=& \sum_{n=1}^{N}
\left( p_+ p_-  +  [N-n]n \right).
\end{eqnarray}
Recall that $\sum_{n=1}^{N} n^2 = N(N+1)(2N+1)/6$. A little work now gives
\begin{equation}
\sum_{n=1}^{N} [N-n]n = {(N-1)N(N+1)\over6} = {N+1\choose3}_.
\end{equation}
Collecting terms and rearranging
\begin{equation}
[G_R Q]\{sI\}
= s_0 \left\{ N p_+ p_-  + {N+1\choose3} \right\} \;
 ( s_0 [q^\bullet{}_\bullet] )_.
\end{equation}

%%%%%%%%%%%%%%%%%%%%%%%%%%%%%%%%%%%%%%%%%%%%%%%%%%%%%%%%%%%%%%%%%%%%%%%%%%%%%%
\subsubsection{Second--order term}
%%%%%%%%%%%%%%%%%%%%%%%%%%%%%%%%%%%%%%%%%%%%%%%%%%%%%%%%%%%%%%%%%%%%%%%%%%%%%%

The second--order term in the expansion is simply
\begin{equation}
{[G_R Q]}^2\{sI\}
= \sum_{n=2}^{N} \sum_{m=1}^{n-1} (s-s_n) (s_n-s_m)  s_m
   \times [q^\bullet{}_\bullet]^2.
\end{equation}
Extracting all dimensionfull factors
\begin{equation}
{[G_R Q]}^2\{sI\}
= s_0 \; S(N,2) \; ( s_0 [q^\bullet{}_\bullet])^2.
\end{equation}
The series summation is a trifle messy:
\begin{eqnarray}
S(N,2)
&=& \sum_{n=2}^{N} \sum_{m=1}^{n-1} (p_- + [N-n]) [n-m] (p_+ + [m-1]),
\nonumber\\
&=& \sum_{n=1}^{N} \sum_{m=1}^n (p_- + [N-n]) [n-m] (p_+ + [m-1]).
\end{eqnarray}
Expanding the sum
\begin{eqnarray}
S(N,2) = \sum_{n=1}^{N} \sum_{m=1}^{n}
&\Bigg\{& p_+ p_- [n-m] \nonumber\\
&+& p_+ ( [N-n] [n-m] ) \nonumber\\
&+& p_- ( [n-m][m-1] ) \nonumber\\
&+&  [N-n] [n-m][m-1] \Bigg\}.
\end{eqnarray}
The terms linear in $p_+$ and $p_-$ have identical coefficients.
They may be combined using $p_+ + p_- = 1$, and can then be collapsed
onto the constant term. This can be seen by noting
\begin{eqnarray}
    \sum_{n=1}^{N} \sum_{m=1}^{n} [N-n] [n-m]
&=& \sum_{m=1}^{N} \sum_{n=m}^N [N-n] [n-m] \nonumber\\
&=& \sum_{m=1}^{N} \sum_{n=1}^{N-m+1} [N-m+1-n] [n-1] \nonumber\\
&=& \sum_{m=1}^{N} \sum_{n=1}^{m} [m-n] [n-1] \nonumber\\
&=& \sum_{n=1}^{N} \sum_{m=1}^{n} [n-m] [m-1].
\end{eqnarray}
This reduces the summation to
\begin{equation}
S(N,2) = \sum_{n=1}^{N} \sum_{m=1}^{n}
\left\{ p_+ p_- [n-m]+  [N-n][n-m] m \right\}.
\end{equation}
Note
\begin{eqnarray}
\sum_{n=1}^{N} \sum_{m=1}^{n} [n-m]
&=& \sum_{n=1}^{N} \left[ n^2 -{n(n+1)\over2} \right]\nonumber\\
&=& \sum_{n=1}^{N} {n(n-1)\over2} \nonumber\\
&=&  {(N-1)N(N+1)\over6} = {N+1\choose3}_.
\end{eqnarray}
To evaluate the remaining term in $S(N,2)$, recall the identities
\begin{equation}
{n+1\choose k+1} = {n\choose k} + {n\choose k+1}_.
\end{equation}
and
\begin{equation}
\sum_{n=1}^N {n\choose k} = \sum_{n=k}^N {n\choose k} = {N+1\choose k+1}_.
\end{equation}
Simple but tedious manipulations now give
\begin{eqnarray}
\sum_{n=1}^{N} \sum_{m=1}^{n} [N-n][n-m] m
&=& \sum_{n=1}^{N} [N-n] {n+1\choose3}
\nonumber\\
&=& \sum_{n=1}^{N} [(N+2) - (n+2)] {n+1\choose3}
\nonumber\\
&=& \sum_{n=1}^{N} \left\{ [N+2] {n+1\choose3} - 4 {n+2\choose4} \right\}
\nonumber\\
&=& [N+2] {N+2\choose4} - 4 {N+3\choose5}
\nonumber\\
&=& {N+2\choose5}_.
\end{eqnarray}
So finally
\begin{equation}
S(N,2) = {N+1\choose3}p_+ p_- +  {N+2\choose5}_.
\end{equation}
The beginning of a pattern might be discerned from this result.

%%%%%%%%%%%%%%%%%%%%%%%%%%%%%%%%%%%%%%%%%%%%%%%%%%%%%%%%%%%%%%%%%%%%%%%%%%%%%%
\subsubsection{n'th--order term}
%%%%%%%%%%%%%%%%%%%%%%%%%%%%%%%%%%%%%%%%%%%%%%%%%%%%%%%%%%%%%%%%%%%%%%%%%%%%%%

The general n'th--order term is
\begin{eqnarray}
{[G_R Q]}^n\{sI\} =&&
  \sum_{m_n=n}^{N}
  \sum_{m_{n-1}=n-1}^{m_n}
  \sum_{m_{n-2}=n-2}^{m_{n-1}}
  \cdots
  \sum_{m_3=3}^{m_4}
  \sum_{m_2=2}^{m_3}
  \sum_{m_1=1}^{m_2}
\nonumber\\
&& \qquad \vphantom{\sum_n^N}
(s-s_{m_n}) (s_{m_n}-s_{m_{n-1}}) \cdots  (s_{m_2}-s_{m_1}) s_{m_1}
\nonumber\\
&& \qquad \times [q^\bullet{}_\bullet]^n.
\end{eqnarray}
Extracting all dimensionfull factors
\begin{equation}
{[G_R Q]}^n\{sI\}
= s_0 \; S(N,n) \; ( s_0 [q^\bullet{}_\bullet])^n.
\end{equation}
The relevant series sum is
\begin{eqnarray}
S(N,n) \equiv&&
  \sum_{m_n=n}^{N}
  \sum_{m_{n-1}=n-1}^{m_n}
  \sum_{m_{n-1}=n-2}^{m_{n-1}}
  \cdots
  \sum_{m_3=3}^{m_4}
  \sum_{m_2=2}^{m_3}
  \sum_{m_1=1}^{m_2}
\nonumber\\
&& \qquad \vphantom{\sum_n^N}
(p_- +[N-{m_n}]) ({m_n}-{m_{n-1}}) \cdots  ({m_2}-{m_1})(p_+ +[m_1-1]).
\end{eqnarray}
This has the form
\begin{equation}
S(N,n) = j(N,n) p_+ p_- + k^+(N,n) p_+ + k^-(N,n) p_- + l(N,n).
\end{equation}
Here
\begin{eqnarray}
 k^+(N,n) \equiv&&
\sum_{m_n=n}^{N}
\sum_{m_{n-1}=m_{m_2}}^{m_n}
\sum_{m_{n-1}=1}^{n-1}
\cdots
\sum_{m_3=3}^{m_4}
\sum_{m_2=2}^{m_3}
\sum_{m_1=1}^{m_2}
\nonumber\\
&&\qquad \vphantom{\sum_n^N}
(N-m_n)({m_n}-{m_{n-1}}) \cdots (m_3 - m_2) ({m_2}-{m_1})
\nonumber\\
=&&
\sum_{m=n}^{N} [N-m] k^+(m, n).
\end{eqnarray}
On the other hand, $k^-(N,n)$ satisfies
\begin{eqnarray}
 k^-(N,n) &\equiv&
\sum_{m_n=n}^{N}
\sum_{m_{n-1}=n-1}^{m_n}
\sum_{m_{n-2}=n-2}^{m_{n-1}}
\cdots
\sum_{m_3=3}^{m_4}
\sum_{m_2=2}^{m_3}
\sum_{m_1=1}^{m_2}
\nonumber\\
&&\qquad \vphantom{\sum_n^N}
(m_n - m_{n-1}) (m_{n-1} - m_{n-2}) \cdots (m_2 - m_1)(m_1 - 1)
\nonumber\\
&=&
\sum_{m_1=1}^{N-n}
\sum_{m_2=m_1}^{N-n+1}
\sum_{m_3=m_2}^{N-n+2}
\cdots
\sum_{m_{n-2}=m_{n-3}}^{N-2}
\sum_{m_{n-1}=m_{n-2}}^{N-1}
\sum_{m_n=m_{n-1}}^{N}
\nonumber\\
&&\qquad \vphantom{\sum_n^N}
(m_n - m_{n-1}) (m_{n-1} - m_{n-2}) \cdots (m_2 - m_1)(m_1 - 1)
\nonumber\\
&=&
\sum_{m=1}^{N-n} k^-(N-m+1,n-1) [m-1]
\nonumber\\
&=&
\sum_{m=n}^{N} [N-m] k^-(m,n-1).
\end{eqnarray}
So $k^+$ and $k^-$ satisfy the same recursion relation. But
\begin{equation}
k^+(N,1) = \sum_{m=1}^N (N-m) = {N(N-1)\over2} =
           \sum_{m=1}^N (m-1) = k^-(N,1).
\end{equation}
This is enough to tell one that for all $(N,n)$, $k^+(N,n) =
k^-(N,n)$.  One can actually do better than this by guessing the
solution to the recursion equation and checking the result
\begin{equation}
k^+(N,n) = k^-(N,n) = {N+n-1\choose2n}_.
\end{equation}
To prove this, note
\begin{eqnarray}
\sum_{m=n}^N [N-m] {m+n-2\choose2n-2}
&=&
\sum_{m=n}^N \left\{ [N+n-1] - [m+n-1] \right\} {m+n-2\choose2n-2}
\nonumber\\
&=&
\sum_{m=n}^N \left\{ [N+n-1] {m+n-2\choose2n-2}
                     - [2n-1]  {m+n-1\choose2n-1} \right\}
\nonumber\\
&=&
[N+n-1] {N+n-1\choose2n-1} - [2n-1] {N+n\choose2n}
\nonumber\\
&=&
{N+n-1\choose2n}_.
\end{eqnarray}
It is now useful to define
\begin{equation}
\ell(N,n) \equiv l(N,n) + k^+(N,n) \equiv l(N,n) + k^-(N,n).
\end{equation}
Explicitly
\begin{eqnarray}
\ell(N,n) &\equiv&
\sum_{m_n=n}^{N}
\sum_{m_{n-1}=n-1}^{m_n}
\sum_{m_{n-2}=n-2}^{m_{n-1}}
\cdots
\sum_{m_3=3}^{m_4}
\sum_{m_2=2}^{m_3}
\sum_{m_1=1}^{m_2}
\nonumber\\
&&\qquad \vphantom{\sum_n^N}
(N-m_n)({m_n}-{m_{n-1}}) \cdots (m_3 - m_2) ({m_2}-{m_1}) m_1
\nonumber\\
&=&
\sum_{m=n}^{N} [N-m] \ell(m,n-1).
\end{eqnarray}
{}From previous computations one already knows $\ell(N,1)={N+1\choose3}$,
and $\ell(N,2)={N+2\choose5}$. Minor modifications of the technique
used in evaluating $k^\pm(N,n)$ shows
\begin{equation}
\ell(N,n) = {N+m\choose 2n+1}_.
\end{equation}
Finally, turn attention to the quantity
\begin{eqnarray}
 j(N,n) &\equiv&
\sum_{m_n=n}^{N}
\sum_{m_{n-1}=n-1}^{m_n}
\sum_{m_{n-2}=n-2}^{m_{n-1}}
\cdots
\sum_{m_3=3}^{m_4}
\sum_{m_2=2}^{m_3}
\sum_{m_1=1}^{m_2}
\nonumber\\
&&\qquad \vphantom{\sum_n^N}
(m_n - m_{n-1}) (m_{n-1} - m_{n-2}) \cdots (m_3 - m_2) (m_2 - m_1)
\nonumber\\
&=& \sum_{m=n}^{N} k^+(m,n-1)
\nonumber\\
&=& \sum_{m=n}^{N} {m+n-2\choose2n-2}
\nonumber\\
&=& {N+n-1\choose2n-1}_.
\end{eqnarray}
Combining everything, one obtains the rather simple and pleasing result
\begin{equation}
S(N,n) = {N+n-1\choose 2n-1} p_+ p_- + {N+n\choose 2n+1}_.
\end{equation}
The result is in fact so simple that one suspects that there should
be easier ways of deriving it. I reiterate the general
n'th order term:
\begin{equation}
{[G_R Q]}^n\{sI\}
= s_0 \left\{ {N+n-1\choose 2n-1} p_+ p_-  + {N+n\choose 2n+1} \right\}
\left( s_0 [q^\bullet{}_\bullet]\right)^n.
\end{equation}

%%%%%%%%%%%%%%%%%%%%%%%%%%%%%%%%%%%%%%%%%%%%%%%%%%%%%%%%%%%%%%%%%%%%%%%%%%%%%%
\subsubsection{(N-1)'th--order term}
%%%%%%%%%%%%%%%%%%%%%%%%%%%%%%%%%%%%%%%%%%%%%%%%%%%%%%%%%%%%%%%%%%%%%%%%%%%%%%

The (N-1)'th term in the expansion is relatively simple to write
down explicitly
\begin{eqnarray}
[G_R Q]^{N-1}\{sI\}
&=& \Bigg\{(s_-+s_0) (s_0)^{N-2} s_+   \nonumber\\
        && +
        \sum_{n=2}^{N-1}  s_-  (s_0)^{N-(n+1)} (2 s_0) (s_0)^{(n-1)-1} s_+
	\nonumber\\
	&&+
	s_- (s_0)^{N-2} (s_+ + s_0)
\Bigg\} [q^{\bullet}{}_\bullet]^{N-1} \nonumber\\
&=& \left\{ 2(N-1) {s_+ s_-\over s_0} + s_0\right\}
( s_0  [q^\bullet{}_\bullet])^{N-1}\nonumber\\
&=& s_0 \left\{ 2(N-1) p_+ p_- + 1\right\}
( s_0  [q^\bullet{}_\bullet])^{N-1}.
\end{eqnarray}
This result serves as a check on the general expression for the
n'th order term.

%%%%%%%%%%%%%%%%%%%%%%%%%%%%%%%%%%%%%%%%%%%%%%%%%%%%%%%%%%%%%%%%%%%%%%%%%%%%%%
\subsubsection{N'th--order term}
%%%%%%%%%%%%%%%%%%%%%%%%%%%%%%%%%%%%%%%%%%%%%%%%%%%%%%%%%%%%%%%%%%%%%%%%%%%%%%

Ultimately
\begin{equation}
[G_R Q]^N\{sI\}
=
s_- {s_0}^{N-1} s_+ [q^{\bullet}{}_{\bullet}]^N
= s_0 \; p_+ p_-  \; ( s_0  [q^\bullet{}_\bullet])^N.
\end{equation}
Again, this serves as a check on the general expression for the
n'th order term.

%%%%%%%%%%%%%%%%%%%%%%%%%%%%%%%%%%%%%%%%%%%%%%%%%%%%%%%%%%%%%%%%%%%%%%%%%%%%%%
\subsubsection{Collecting terms}
%%%%%%%%%%%%%%%%%%%%%%%%%%%%%%%%%%%%%%%%%%%%%%%%%%%%%%%%%%%%%%%%%%%%%%%%%%%%%%

One is now in a position to write down a closed form expression
for the van Vleck determinant
\begin{eqnarray}
\Delta_\gamma^{-1}
&=& \det\left[ s_\gamma(x,y)^{-1} A(s) \right]
\nonumber\\
&=&
\det\left[  s^{-1}
        \left(I + [G_R Q] + [G_R Q]^2 + \cdots + [G_R Q]^N \right) \{sI\}
	\right]
\nonumber\\
&=&
\det\left[ \sum_{n=0}^N  S(N,n)
( s_0  [q^\bullet{}_\bullet])^n /N \right]
\nonumber\\
&=&\det\Bigg[
\left\{[I^\bullet{}_\bullet] +
        p_+ p_- ( s_0 [q^\bullet{}_\bullet]) \right\} \times
\nonumber\\
&& \qquad\left\{ \sum_{n=0}^{N-1} {N+n\choose 2n+1}
{(s_0 [q^\bullet{}_\bullet])^n  \over N} \right\} \Bigg]
\nonumber\\
&=& \det\left\{ [I^\bullet{}_\bullet] +
                 p_+ p_- (s_0[q^\bullet{}_\bullet])\right\} \times
\nonumber\\
&&\qquad \det\left\{ \sum_{n=0}^{N-1} {N+n\choose 2n+1}
          {(s_0 [q^\bullet{}_\bullet])^n \over N} \right\}_.
\end{eqnarray}
This factorization property for the determinant is a quite remarkable
and unexpected feature. Equally remarkable is the fact that the
polynomials occurring in the above expression,
\begin{equation}
P_N(z) = \sum_{n=0}^{N} {N+n+1\choose 2n+1} \; z^n,
\end{equation}
are nothing more than suitably disguised Chebyshev polynomials
({\em a.k.a.} Tschebischeff polynomials) of the second
kind~\cite{G-and-R,A-and-S,Spiegel}. Indeed
\begin{equation}
P_N(z) = U_N([z+4]/2).
\end{equation}
(Since I will not need to use this result, I leave the proof as an
exercise for the reader.)

Now, provided there is no accidental zero suppressing the highest
order term, one may approximate
\begin{eqnarray}
\Delta_\gamma^{-1}
\approx
\det\left\{ [I^\bullet{}_\bullet]+ p_+p_- (s_0[q^\bullet{}_\bullet])\right\}
{\det}'\left\{ {(s_0 [q^\bullet{}_\bullet])^{N-1} \over N} \right\}_.
\end{eqnarray}
Far away from the mouth of the wormhole $p_+ p_- \gg 0$, so that
\begin{equation}
\Delta_\gamma
\approx
\left({ N s_0^2 \over s_+ s_- }\right)^{\#}
{\det}'\left\{s_0  [q^\bullet{}_\bullet] \right\}^{-N}.
\end{equation}
This should be compared to eq. (64) of Kim and Thorne~\cite{Kim-Thorne}.
If the base point of the geodesic lies on the throat of the wormhole
$p_+ p_- = 0$. The dominant term in the determinant is now
\begin{equation}
\Delta_\gamma
\approx {N}^\# \; {\det}'\left\{s_0  [q^\bullet{}_\bullet] \right\}^{-N+1}.
\end{equation}
Compare with eq (48) of Kim and Thorne~\cite{Kim-Thorne}. A benefit
of the current analysis is that it is now possible to probe the
transition region as the base point of the geodesic approaches the
throat of the wormhole. ({\em i.e.} $s_+s_- \to 0$).

Further simplifications require more precise model building for
the wormhole in question. Useful models may be obtained by extending
the usual thin--shell formalism ({\em a.k.a.} the junction condition
formalism).

%%%%%%%%%%%%%%%%%%%%%%%%%%%%%%%%%%%%%%%%%%%%%%%%%%%%%%%%%%%%%%%%%%%%%%%%%%%%%%
\section{EXTENDED THIN SHELL FORMALISM: THEORY}
%%%%%%%%%%%%%%%%%%%%%%%%%%%%%%%%%%%%%%%%%%%%%%%%%%%%%%%%%%%%%%%%%%%%%%%%%%%%%%

Development of the junction condition formalism, initiated by
Lanczos~\cite{Lanczos} and Sen~\cite{Sen} in the early part of this
century, culminated in the work of Israel~\cite{Israel}, of Taub
\cite{Taub}, and of Barrabes~\cite{Barrabes}.  The Lanczos--Sen--Israel
version of the formalism relates the Ricci tensor generated by the
gravitational influence a thin shell of stress--energy to the
discontinuity of the second fundamental forms defined by considering
the thin shell to be an immersed submanifold of the full spacetime.
Unfortunately, for the present application a knowledge of the Ricci
tensor is not enough.  An expression for the full Riemann tensor
is necessary.

For the purposes of this paper,  it is sufficient to consider a
thin shell of stress--energy residing in an otherwise flat Lorentzian
spacetime.  Such a system is described by a number of copies of
Minkowski space with two or more regions excised, and with the
boundaries of those regions identified in some suitable manner.
Examples of such systems include the traversable wormholes of
interest in this paper (in the limit as the throat of the wormhole
becomes arbitrarily short), and the cellular cosmologies of Redmount
\cite{Redmount}. The Riemann tensor for such a geometry is identically
zero everywhere except at the boundary  where the identification
procedure takes place.  The formalism is  presented for a timelike
shell (spacelike normal). The modifications required to deal with
a spacelike shell (timelike normal) are trivial.  The general
analysis for the case  of a shell of null stress--energy is
considerably more tedious, and will not be needed in this paper.
The reader is referred to the work of Redmount~\cite{Redmount-Null}
where an analysis in terms of the Newman--Penrose formalism is
presented.

The modifications required in order to permit the wormhole to reside
in a non--flat background are sufficiently simple that, in the
interests of generality, the appropriate modifications are
included.

%%%%%%%%%%%%%%%%%%%%%%%%%%%%%%%%%%%%%%%%%%%%%%%%%%%%%%%%%%%%%%%%%%%%%%%%%%%%%%
\subsection{Second fundamental form}
%%%%%%%%%%%%%%%%%%%%%%%%%%%%%%%%%%%%%%%%%%%%%%%%%%%%%%%%%%%%%%%%%%%%%%%%%%%%%%

Consider a thin shell of stress--energy situated  in some smooth
(possibly curved) ambient space.  Adopt Gaussian normal coordinates
near the shell.  That is: $\eta=0$ is the location of the shell,
while $\eta$ itself is a normal coordinate.  Without loss of
generality the metric can, in the vicinity of the shell, be cast
in the form
\begin{equation}
g_{\alpha\beta}(\eta,x_\perp)
= \Theta(\eta)  \; g^+_{\alpha\beta}(\eta,x_\perp)
+ \Theta(-\eta) \; g^-_{\alpha\beta}(\eta,x_\perp).
\end{equation}
The metric itself is continuous at the shell so that
\begin{equation}
g^+_{\alpha\beta}(\eta=0,x_\perp) =
g^-_{\alpha\beta}(\eta=0,x_\perp).
\end{equation}
The second fundamental forms associated with the two sides of the
shell $\eta=0$ are simply
\begin{eqnarray}
K^+_{\alpha\beta}
&= {1\over2} \left.{\partial g_{\alpha\beta}\over\partial\eta} \right|_{0^+}
&= {1\over2} {\partial g^+_{\alpha\beta}\over\partial\eta}, \\
K^-_{\alpha\beta}
&= {1\over2} \left.{\partial g_{\alpha\beta}\over\partial\eta} \right|_{0^-}
&= {1\over2} {\partial g^-_{\alpha\beta}\over\partial\eta}.
\end{eqnarray}
The discontinuity in the second fundamental form is
\begin{equation}
\kappa_{\alpha\beta}
= K^+_{\alpha\beta} - K^-_{\alpha\beta},
\end{equation}
while the normal to the thin shell satisfies
\begin{equation}
n_\alpha = \partial_\alpha \eta; \qquad
n^\alpha n_\alpha = +1; \qquad
n^\alpha \; \kappa_{\alpha\beta} =0.
\end{equation}

Consider the metric derivatives $g_{\alpha\beta,\gamma}$. In view
of the fact that ${d\over dx} \Theta(\pm\eta) = \pm \delta(\eta)$,
and of the continuity of the metric at $\eta=0$, one derives
\begin{equation}
g_{\alpha\beta,\gamma}
= \Theta(\eta)  \; g^+_{\alpha\beta,\gamma}
+ \Theta(-\eta) \; g^-_{\alpha\beta,\gamma}.
\end{equation}
It is now a simple application of Gaussian normal coordinates to
show~\cite{Lanczos}
\begin{equation}
 g^+_{\alpha\beta,\gamma} -
 g^-_{\alpha\beta,\gamma} =
 2 \kappa_{\alpha\beta} \; n_\gamma.
\end{equation}
Though most easily derived using Gaussian normal coordinates this
result is general. The second derivatives of the metric are easily
evaluated
\begin{eqnarray}
g_{\alpha\beta,\gamma\delta}
&&= \Theta(\eta)  \; g^+_{\alpha\beta,\gamma\delta}
  + \Theta(-\eta) \; g^-_{\alpha\beta,\gamma\delta}
  + \delta(\eta) \left\{  g^+_{\alpha\beta,\gamma} -
    g^-_{\alpha\beta,\gamma} \right\} n_\delta, \\
&&= \Theta(\eta)  \; g^+_{\alpha\beta,\gamma\delta}
  + \Theta(-\eta) \; g^-_{\alpha\beta,\gamma\delta}
  + 2 \delta(\eta) \; \kappa_{\alpha\beta} \; n_\gamma n_\delta.
\end{eqnarray}
In a very similar vein, one may consider the Christoffel symbol
\begin{equation}
\Gamma_{\alpha\beta\gamma}
\equiv {1\over2} \left[ g_{\alpha\beta,\gamma}
                        + g_{\alpha\gamma,\beta}
			- g_{\beta\gamma,\alpha} \right] =
   \Theta(\eta)  \; \Gamma^+_{\alpha\beta\gamma}
+  \Theta(-\eta) \; \Gamma^-_{\alpha\beta\gamma}.
\end{equation}

%%%%%%%%%%%%%%%%%%%%%%%%%%%%%%%%%%%%%%%%%%%%%%%%%%%%%%%%%%%%%%%%%%%%%%%%%%%%%%
\subsection{Riemann tensor}
%%%%%%%%%%%%%%%%%%%%%%%%%%%%%%%%%%%%%%%%%%%%%%%%%%%%%%%%%%%%%%%%%%%%%%%%%%%%%%

With all of the preparatory work out of the way, it is now a simple
matter of invoking the standard definition of the Riemann tensor
\begin{eqnarray}
R_{\alpha\beta\gamma\delta} =
&-&{ 1\over2}
\left(g_{\alpha\gamma,\beta\delta} + g_{\beta\delta,\alpha\gamma}
   - g_{\alpha\delta,\beta\gamma} - g_{\beta\gamma,\alpha\delta} \right)
\nonumber\\
&-& g^{\sigma\rho}
\left( \Gamma_{\alpha\gamma\sigma} \Gamma_{\beta\delta\rho}
- \Gamma_{\alpha\delta\sigma} \Gamma_{\beta\gamma\rho} \right),
\end{eqnarray}
to see that the Riemann tensor in the vicinity of the thin shell
is of the form
\begin{eqnarray}
R_{\alpha\beta\gamma\delta}
&=& - \delta(\eta)
\left[ \kappa_{\alpha\gamma} \; n_\beta  \; n_\delta
     + \kappa_{\beta\delta}  \; n_\alpha \; n_\gamma
     - \kappa_{\alpha\delta} \; n_\beta  \; n_\gamma
     - \kappa_{\beta\gamma}  \; n_\alpha \; n_\delta  \right]
\nonumber \\
&&\qquad + \Theta(\eta)  \; R^+_{\alpha\beta\gamma\delta}
         + \Theta(-\eta) \; R^-_{\alpha\beta\gamma\delta}.
\end{eqnarray}
It is easy to see that the known symmetries of the Riemann tensor
are correctly reflected in this expression. A little more work
establishes that the full (uncontracted) Bianchi identities are
satisfied. To check this, note that $\partial_\alpha \delta(\eta)
= \delta'(\eta) \partial_\alpha \eta = \delta'(\eta) n_\alpha$,
and that $\nabla_\alpha n_\beta = K^\pm_{\alpha\beta}$ is symmetric.
The proof of the Bianchi identities follows by considering the
antisymmetrization properties of $R_{\alpha\beta[\gamma\delta;\epsilon]}$.

With hindsight, the above expression for the full Riemann tensor
can be easily derived from standard textbook results. For instance,
from equations (21.82), (21.75), and (21.76) of Misner, Thorne,
and Wheeler~\cite{MTW} a gaussian pillbox integration similar to
that discussed on pages 551 through 556 rapidly leads to the above
result.

By contraction on appropriate indices one may recover the known
Lanczos--Sen--Israel junction
conditions~\cite{Lanczos,Sen,Israel,Taub,Barrabes}.  In terms of
the Ricci tensor, the Einstein tensor, and the Ricci scalar:
\begin{eqnarray}
R_{\alpha\beta} &= &- \delta(\eta)
\left[ \kappa_{\alpha\beta} + \kappa \; n_\alpha n_\beta \right]
+  \Theta(\eta) \; R^+_{\alpha\beta} + \Theta(-\eta) \; R^-_{\alpha\beta}.
\\
R &=& - 2 \kappa \; \delta(\eta)
+  \Theta(\eta) \; R^+ + \Theta(-\eta) \; R^-.
\\
E_{\alpha\beta} &=& - \delta(\eta)
\left[ \kappa_{\alpha\beta}
     - \kappa ( g_{\alpha\beta} - n_\alpha n_\beta )\right]
+ \Theta(\eta) \; E^+_{\alpha\beta} + \Theta(-\eta) \; E^-_{\alpha\beta}.
\end{eqnarray}
These equations are completely equivalent to the usual Lanczos--Sen--Israel
version of the junction condition formalism. Indeed, this is a
quick consistency check.

%%%%%%%%%%%%%%%%%%%%%%%%%%%%%%%%%%%%%%%%%%%%%%%%%%%%%%%%%%%%%%%%%%%%%%%%%%%%%%
\subsection{Jacobi tensor}
%%%%%%%%%%%%%%%%%%%%%%%%%%%%%%%%%%%%%%%%%%%%%%%%%%%%%%%%%%%%%%%%%%%%%%%%%%%%%%

Tidal effects are sometimes reformulated in terms of the Jacobi
tensor~\cite{MTW}
\begin{eqnarray}
J_{\alpha\beta\gamma\delta} &\equiv&
{1\over2} \left( R_{\alpha\gamma\beta\delta}
               + R_{\alpha\delta\beta\gamma} \right) \\
R_{\alpha\beta\gamma\delta} &\equiv&
{2\over3} \left( J_{\alpha\gamma\beta\delta}
               + J_{\alpha\delta\beta\gamma} \right)
\end{eqnarray}
This easily leads to
\begin{eqnarray}
J_{\alpha\beta\gamma\delta}
&=& - \delta(\eta)
\Big[      \kappa_{\alpha\beta}  \; n_\gamma \; n_\delta
     +     \kappa_{\gamma\delta} \; n_\alpha \; n_\beta
           \nonumber\\
&&\qquad - {1\over2} \kappa_{\alpha\delta} \; n_\beta  \; n_\gamma
         - {1\over2} \kappa_{\beta\gamma}  \; n_\alpha \; n_\delta
         - {1\over2} \kappa_{\alpha\gamma} \; n_\beta  \; n_\delta
         - {1\over2} \kappa_{\beta\delta}  \; n_\alpha \; n_\gamma  \Big]
\nonumber \\
&&\qquad + \Theta(\eta)  \; J^+_{\alpha\beta\gamma\delta}
         + \Theta(-\eta) \; J^-_{\alpha\beta\gamma\delta}.
\end{eqnarray}
This result is, of course, completely equivalent to the result for
the Riemann tensor.

%%%%%%%%%%%%%%%%%%%%%%%%%%%%%%%%%%%%%%%%%%%%%%%%%%%%%%%%%%%%%%%%%%%%%%%%%%%%%%
\subsection{Weyl tensor}
%%%%%%%%%%%%%%%%%%%%%%%%%%%%%%%%%%%%%%%%%%%%%%%%%%%%%%%%%%%%%%%%%%%%%%%%%%%%%%

The Weyl tensor describes the part of the curvature that is invariant
under conformal rescalings of the metric. It may be computed by
brute force starting from the definition
\cite{MTW,Hawking-Ellis}
\begin{eqnarray}
W_{\alpha\beta\gamma\delta} \equiv
&+& R_{\alpha\beta\gamma\delta} \nonumber\\
&-& {1\over2}
\left[ g_{\alpha\gamma} \; R_{\beta\delta}
     + g_{\beta\delta}  \; R_{\alpha\gamma}
     - g_{\alpha\delta} \; R_{\beta\gamma}
     - g_{\beta\gamma}  \; R_{\alpha\delta}  \right]
     \nonumber \\
&+&{1\over12} R
\left[ g_{\alpha\gamma} \; g_{\beta\delta}
     + g_{\beta\delta}  \; g_{\alpha\gamma}
     - g_{\alpha\delta} \; g_{\beta\gamma}
     - g_{\beta\gamma}  \; g_{\alpha\delta}  \right].
\end{eqnarray}
To simplify life, define the transverse metric to be $h_{\alpha\beta}
\equiv g_{\alpha\beta} - n_\alpha n_\beta$, and define the transverse
traceless part of the discontinuity in the second fundamental form
to be ${\tilde\kappa}_{\alpha\beta} \equiv \kappa_{\alpha\beta} -
{1\over3} \kappa h_{\alpha\beta}$.  The result of a little tedious
algebra gives
\begin{eqnarray}
W_{\alpha\beta\gamma\delta}
&=& - \delta(\eta)
\Big[ {\tilde\kappa}_{\alpha\gamma} \;
              (n_\beta  \; n_\delta - {1\over2} g_{\beta\delta} )
     + {\tilde\kappa}_{\beta\delta}  \;
              (n_\alpha \; n_\gamma - {1\over2} g_{\alpha\gamma})
     \nonumber\\
&&\qquad - {\tilde\kappa}_{\alpha\delta} \;
              (n_\beta  \; n_\gamma - {1\over2} g_{\beta\gamma} )
         - {\tilde\kappa}_{\beta\gamma}  \;
              (n_\alpha \; n_\delta - {1\over2} g_{\alpha\delta}) \Big]
\nonumber \\
&&\qquad + \Theta(\eta)  \; W^+_{\alpha\beta\gamma\delta}
         + \Theta(-\eta) \; W^-_{\alpha\beta\gamma\delta},
\end{eqnarray}
which is easily verified to satisfy the relevant trace identities.
As an aside, it is relatively easy to convince oneself that the
tensors  ${\tilde\kappa}^\alpha{}_\beta \equiv \kappa^\alpha{}_\beta
- {1\over3} \kappa h^\alpha{}_\beta$, and $n^\alpha n_\beta$ are
separately invariant under conformal transformations of the metric.

%%%%%%%%%%%%%%%%%%%%%%%%%%%%%%%%%%%%%%%%%%%%%%%%%%%%%%%%%%%%%%%%%%%%%%%%%%%%%%
\subsection{Examples}
%%%%%%%%%%%%%%%%%%%%%%%%%%%%%%%%%%%%%%%%%%%%%%%%%%%%%%%%%%%%%%%%%%%%%%%%%%%%%%

In the idealised case of a traversable wormhole constructed from
flat Minkowski space by cut and paste techniques~\cite{Visser89a,Visser89b}
one  knows that the Riemann tensor must be zero everywhere except
at the throat itself. So in this idealised case one obtains the
exact result
\begin{equation}
R_{\alpha\beta\gamma\delta} = - \delta(\eta)
\left[ \kappa_{\alpha\gamma} \; n_\beta  \; n_\delta
     + \kappa_{\beta\delta}  \; n_\alpha \; n_\gamma
     - \kappa_{\alpha\delta} \; n_\beta  \; n_\gamma
     - \kappa_{\beta\gamma}  \; n_\alpha \; n_\delta  \right]
\end{equation}
Similar considerations apply to Redmount's cellular cosmologies
\cite{Redmount}. In both cases one now has a very powerful tool
for attacking problems involving tidal effects.

%%%%%%%%%%%%%%%%%%%%%%%%%%%%%%%%%%%%%%%%%%%%%%%%%%%%%%%%%%%%%%%%%%%%%%%%%%%%%%
\section{EXTENDED THIN SHELL FORMALISM: APPLICATION}
%%%%%%%%%%%%%%%%%%%%%%%%%%%%%%%%%%%%%%%%%%%%%%%%%%%%%%%%%%%%%%%%%%%%%%%%%%%%%%

%%%%%%%%%%%%%%%%%%%%%%%%%%%%%%%%%%%%%%%%%%%%%%%%%%%%%%%%%%%%%%%%%%%%%%%%%%%%%%
\subsection{Riemann tensor}
%%%%%%%%%%%%%%%%%%%%%%%%%%%%%%%%%%%%%%%%%%%%%%%%%%%%%%%%%%%%%%%%%%%%%%%%%%%%%%

It is possible to improve and make more explicit the general analysis
of the van Vleck determinant by applying the extended thin--shell
formalism as outlined above.  For a wormhole in the short--throat
flat--space approximation the full Riemann tensor is
\begin{equation}
R_{\alpha\beta\gamma\delta} = - \delta(\eta)
\left[ \kappa_{\alpha\gamma} \; n_\beta  \; n_\delta
     + \kappa_{\beta\delta}  \; n_\alpha \; n_\gamma
     - \kappa_{\alpha\delta} \; n_\beta  \; n_\gamma
     - \kappa_{\beta\gamma}  \; n_\alpha \; n_\delta  \right]_.
\end{equation}

For the delta function one can easily show
\begin{equation}
\delta(\eta)
= \left(\partial\eta/\partial s\right)^{-1} \delta(s)
= (t \cdot n)^{-1} \; \delta(s).
\end{equation}
The relevant source term driving the tidal evolution equation is
\begin{eqnarray}
Q_{\mu\nu}(s)
&\equiv&
- R_{\mu\alpha\nu\beta} t^\alpha t^\beta \nonumber \\
&=& \delta(\eta)
\left[ \kappa_{\mu\nu} (t\cdot n)^2
     +  n_\mu n_\nu (\kappa_{\alpha\beta} t^\alpha t^\beta)
     - \{(\kappa_{\mu\alpha} t^\alpha)  n_\nu
     + (\kappa_{\nu\alpha} t^\alpha)  n_\mu \}(t \cdot n)  \right] \\
&=& \delta(s)
\left[ \kappa_{\mu\nu} (t\cdot n)
     +  n_\mu n_\nu (\kappa_{\alpha\beta} t^\alpha t^\beta) (t\cdot n)^{-1}
     - \{(\kappa_{\mu\alpha} t^\alpha)  n_\nu
     + (\kappa_{\nu\alpha} t^\alpha)  n_\mu \}  \right]_.
\end{eqnarray}
The discontinuity in the second fundamental form, $\kappa_{\mu\nu}$,
is essentially a measure of the curvature of the wormhole mouth.
Diagonalizing $\kappa$ yields $\kappa_{\mu\nu}= 2\;{\rm diag} \{A,
(1/R_1), (1/R_2), 0\}_{\mu\nu}$. Here $A$ is the
four--acceleration of the wormhole throat --- essentially the radius
of curvature in the timelike direction~\cite{Visser89b}.  $R_{(1,2)}$
are the usual principal radii of curvature in the 3 dimensional
sense. The zero eigenvalue for $\kappa$ reflects the fact that
$\kappa_{\mu\nu} n^\nu = 0$, the second fundamental form is by
construction orthogonal to the normal.

One is now in a position to enunciate a general qualitative result:
\hfil\break (small radius of curvature/large acceleration)
$\Rightarrow$ (large $\kappa$) $\Rightarrow$ (large Riemann tensor)
$\Rightarrow$ (large $A(s)$) $\Rightarrow$ (small van Vleck
determinant).

To proceed further, suppose that the individual wormhole mouths
(though not the full spacetime) are static. This just means that
the individual wormhole mouths are not undergoing  any changes in
internal structure; and also implies that one can safely attach
the notion of a constant four-velocity $V^\mu$ to each individual
wormhole mouth. In particular since each wormhole mouth is now
invariant under time translations along the $V^\mu$ axis one has
$\kappa_{\mu\nu} V^\nu =0$. Further, by construction,  $\kappa_{\mu\nu}
n^\nu =0$ and $V^\mu n_\mu =0$. Indeed in this case $\kappa_{\mu\nu}=
{\rm diag} \{0, (2/R_1), (2/R_2), 0\}_{\mu\nu}$. A brief computation
shows $Q_{\mu\nu} V^\nu =0$. The general analysis already implies
$Q_{\mu\nu} t^\nu =0$.  Therefore, in this approximation two of
the eigenvalues of $Q(s)$ are trivial, and the computation of the
van Vleck determinant reduces to that of a (still decidedly
non--trivial) $2\times2$ matrix.

%%%%%%%%%%%%%%%%%%%%%%%%%%%%%%%%%%%%%%%%%%%%%%%%%%%%%%%%%%%%%%%%%%%%%%%%%%%%%%
\subsection{Single pass: exact result}
%%%%%%%%%%%%%%%%%%%%%%%%%%%%%%%%%%%%%%%%%%%%%%%%%%%%%%%%%%%%%%%%%%%%%%%%%%%%%%

Consider a geodesic that starts at the point $x$, wraps through
the wormhole once, and returns to the point $x$. This geodesic has
two tangent vectors, $t_\pm$, one pointing toward each of the two
wormhole mouths. Each of the two wormhole mouths is characterized
by a four-velocity $V_\pm$, and, at the point where the geodesic
impacts the mouth of the wormhole, a normal $n_\pm$. In view of
the structure outlined above, one can define quantities
$\gamma=(1-\beta^2)^{-1/2}$ and $\theta$ by
\begin{equation}
t_\pm^\mu = \gamma\left(
\beta \; V_\pm^\mu +
\cos\theta \; n_\pm^\mu +
\sin\theta \; z_\pm^\mu
\right).
\end{equation}
Here one has utilized the fact that the tangent vector is (for
current purposes) a spacelike unit vector, and defined another
spacelike unit vector $z$, that lies in the plane of the wormhole
throat.  Because the geodesic must, by construction, pass through
the wormhole mouth in a smooth manner one must have the same
coefficients $\gamma$, $\beta$, and $\theta$, occurring in each of
these two equations.  In particular, $t_+ \cdot n_+ = t_- \cdot
n_- = \gamma \cos\theta$. One may now define the quantity $\ell_\pm
\equiv  \gamma s_\pm = s_\pm \; (t_\pm \cdot n_\pm)/\cos\theta$.
This represents the physical distance between the point $x$ and
the relevant wormhole mouth, as measured in the rest frame of that
wormhole mouth. By extension, one defines $\ell = \ell_+ + \ell_-
= s \, (t_\pm \cdot n_\pm)/\cos\theta$. To facilitate comparison
with reference~\cite{Visser-CPC} one may define ``time shifts'' by
$T_\pm \equiv s_\pm \; (t_\pm \cdot V_\pm)$. Further, one may define
$T=T_+ + T_-$. Finally, as a consistency check, observe $\delta(\eta_+)
= (t_+ \cdot n_+)^{-1} \; \delta(s) = (t_- \cdot n_-)^{-1} \;
\delta(s) = \delta(\eta_-)$.

The van Vleck determinant is simply
\begin{equation}
\Delta_\gamma(x,x)
= \det\left( \delta^\mu{}_\nu
            + {s_+ s_-\over s} [q^\mu{}_\nu] \right)^{-1}_.
\end{equation}
As a result of all the preceding definitions and calculations,
one extracts the relatively compact expression
\begin{eqnarray}
[q^\mu{}_\nu]
&=&
\left[ \kappa_{\mu\nu} (t\cdot n)
     +  n_\mu n_\nu (\kappa_{\alpha\beta} t^\alpha t^\beta) (t\cdot n)^{-1}
     - \{(\kappa_{\mu\alpha} t^\alpha)  n_\nu
     + (\kappa_{\nu\alpha} t^\alpha)  n_\mu \}  \right],
\nonumber\\
&=&
\gamma \cos\theta
\left[ \kappa_{\mu\nu}
     +  n_\mu n_\nu \tan^2\theta (\kappa_{\alpha\beta} z^\alpha z^\beta)
     - \tan\theta \{(\kappa_{\mu\alpha} z^\alpha)  n_\nu
     + (\kappa_{\nu\alpha} z^\alpha)  n_\mu \}  \right].
\end{eqnarray}
Unfortunately this expression, while exact and general, is still
too algebraically messy to be tractable.

%%%%%%%%%%%%%%%%%%%%%%%%%%%%%%%%%%%%%%%%%%%%%%%%%%%%%%%%%%%%%%%%%%%%%%%%%%%%%%
\subsubsection{Radial impact}
%%%%%%%%%%%%%%%%%%%%%%%%%%%%%%%%%%%%%%%%%%%%%%%%%%%%%%%%%%%%%%%%%%%%%%%%%%%%%%

Utilizing the machinery defined above one can define the notion of
a geodesic that radially impacts on the mouth of the wormhole. A
radially impacting geodesic is simply one for which $\theta=0$, so
that one has
\begin{equation}
t_\pm^\mu = \gamma\left( \beta \; V_\pm^\mu + n_\pm^\mu  \right).
\end{equation}
For instance: For spherically symmetric wormhole mouths that directly
face one another, go to any Lorentz frame where the mouths have no
transverse velocity, only longitudinal velocities. In any such
frame pick any point on the line joining the two wormhole mouths.
The geodesic from any such point to itself will impact the wormhole
mouths radially in the sense described above. This is essentially
the notion of the ``central geodesic'' as described by Kim and
Thorne~\cite{Kim-Thorne}.  Returning to the general case, the virtue
of radially impacting geodesics is that for such geodesics one has
the great simplification $\kappa_{\mu\nu} t^\mu =0$, which implies
\begin{equation}
Q_{\mu\nu} \equiv
- R_{\mu\alpha\nu\beta} t^\alpha t^\beta
= \delta(\eta)  \; (t\cdot n)^2 \; \kappa_{\mu\nu}
= \delta(s)  \; (t\cdot n) \; \kappa_{\mu\nu}.
\end{equation}

Let me (temporarily) further restrict the analysis. For a spherically
symmetric wormhole of radius $R$, adopting the time shift identification
of Kim and Thorne~\cite{Kim-Thorne}, the discontinuity in the second
fundamental forms is:
\begin{equation}
\kappa_{\mu\nu} =
{2\over R} \; g^\perp_{\mu\nu} =
{2\over R} \; (g_{\mu\nu} + V_\mu V_\nu - n_\mu n_\nu ).
\end{equation}
Inserting all of this machinery into the previously derived closed
form expression for the single pass van Vleck determinant
\begin{eqnarray}
\Delta_\gamma(s)
&=& \det\left( \delta^\mu{}_\nu
            + {s_+ s_-\over s}
	      {2(t\cdot n)\over R} [g^\perp]^\mu{}_\nu \right)^{-1}_,
	      \nonumber\\
&=& \left( 1 + {s_+ s_-\over s} {2(t \cdot n)\over R}\right)^{-2}_,
              \nonumber\\
&=& \left( 1 + {\ell_+\ell_-\over \ell} {2\over R}\right)^{-2}_.
\end{eqnarray}
This formula nicely interpolates between the near field and far
field results of Kim and Thorne~\cite{Kim-Thorne}.  For points on
the wormhole throat, compare this result to the $N=1$ case of Kim
and Thorne's equation (48).  For points far from the throat compare
this result with the $N=1$ case of their equation (64). Emphasis
should be placed on the fact that the relative velocity of the
wormhole mouths is completely arbitrary. Likewise, the position of
the point $x=y$, though constrained by the condition of radial
impact is otherwise arbitrary.

A simple generalization is to note that there is now nothing sacred
about a spherical wormhole mouth. As long as the geodesic is radially
impacting one may write
\begin{eqnarray}
\Delta_\gamma(x,x)
&=& \det\left( \delta^\mu{}_\nu
            + {s_+ s_-\over s}
	      {\gamma} \kappa^\mu{}_\nu \right)^{-1}_,
	      \nonumber\\
&=& \det\left(  \delta^\mu{}_\nu
        + {\ell_+\ell_-\over \ell} \kappa^\mu{}_\nu \right)^{-1}_,
    \nonumber\\
&=& \left( 1 + {\ell_+\ell_-\over \ell} {2\over R_1}\right)^{-1}
  \left( 1 + {\ell_+\ell_-\over \ell} {2\over R_2}\right)^{-1}_.
\end{eqnarray}
Here, as previously, $R_{(1,2)}$ refer to the two principal radii
of curvature, evaluated at the point where the geodesic impacts
the wormhole mouth. Indeed, consider the cubical (or even polyhedral)
wormholes of reference~\cite{Visser89a}. If  the geodesic impacts
radially on one of the flat faces, $R_{(1,2)}=\infty$, consequently
$\Delta_\gamma(s)=1$.

If the base point $x$ is not on the throat of the wormhole then
$\ell_+\ell_- \neq 0$. If the geodesic impacts radially on one of
the edges, then  one of the principal radii of curvature $R_{(1,2)}$
is zero, (the other is infinite). Consequently $\Delta_\gamma(s)=0$.

If the base point $x$ is on the throat of the wormhole then
$\ell_+\ell_- = 0$. In this case, if the geodesic impacts radially
on one of the edges, one is faced with making sense of the
indeterminate form $0/0$. Proceed as follows: The polyhedral
wormholes of reference~\cite{Visser89a} were constructed by taking
smooth cut--and--paste (thin throat) wormholes and considering the
limit as one of the radii of curvature tends to zero.  For the
issue of interest, the indeterminate $0/0$ is resolved by observing
that one should let the base point  $x$  approach the wormhole
mouth {\em before} letting the wormhole mouth acquire sharp edges
by letting $R_{(1,2)} \to 0$. Consequently $\Delta_\gamma(s)=1$.

Another way of phrasing this is as follows: Letting the wormhole
mouth acquire a sharp edge by letting $R_2 \to 0$ is a somewhat
dubious proposition once $R_2 < \ell_P$.  Radii of curvature smaller
than the Planck length are of doubtful operational significance.
Accepting that the struts supporting edges of a polyhedral wormhole
have thickness of order the Planck length
\begin{equation}
\Delta_\gamma(x,x)^{\rm polyhedral}_{\rm edge-impact}
\approx
\left( 1 + {2\ell_+\ell_-\over \ell \; \ell_P}\right)^{-1}_.
\end{equation}
This ``regulated'' determinant is suitably well behaved as one
approaches the throat.

%%%%%%%%%%%%%%%%%%%%%%%%%%%%%%%%%%%%%%%%%%%%%%%%%%%%%%%%%%%%%%%%%%%%%%%%%%%%%%
\subsubsection{Generic impact}
%%%%%%%%%%%%%%%%%%%%%%%%%%%%%%%%%%%%%%%%%%%%%%%%%%%%%%%%%%%%%%%%%%%%%%%%%%%%%%

To push the analysis beyond the case of radial impact is computationally
messy. It is useful, in the interests of keeping the algebra from
getting too unwieldy, to return to the case of spherical symmetry.
Combining spherical symmetry with a generic impact angle
\begin{eqnarray}
Q_{\mu\nu}(s)
&=& \delta(s)
\left[ \kappa_{\mu\nu} (t\cdot n)
     +  n_\mu n_\nu (\kappa_{\alpha\beta} t^\alpha t^\beta) (t\cdot n)^{-1}
     - \{(\kappa_{\mu\alpha} t^\alpha)  n_\nu
     + (\kappa_{\nu\alpha} t^\alpha)  n_\mu \}  \right] \nonumber\\
&=& \delta(s) {2\gamma\cos\theta\over R}
\left[ g^\perp_{\mu\nu}
     +  n_\mu n_\nu  \tan^2\theta
     - \tan\theta\{z_\mu n_\nu
     + z_\nu n_\mu \}  \right]_.
\end{eqnarray}
One is faced with the task of determining the eigenvalues of the matrix
\begin{equation}
\left[\begin{array}{cccc}
0&0&0&0\\
0&\;\tan^2\theta\;&\;-\tan\theta\;&0\\
0&-\tan\theta&1&0\\
0&0&0&1
\end{array}\right]
\end{equation}
These eigenvalues are easily determined to be $(0;0;1;1+\tan^2\theta)
\equiv (0;0;1;\sec^2\theta)$. Consequently
\begin{equation}
\Delta_\gamma(s)
= \left( 1 + {\ell_+\ell_-\over \ell} {2\over R}\right)^{-1}
  \left( 1 + {\ell_+\ell_-\over \ell} {2\sec^2\theta\over R}\right)^{-1}_.
\end{equation}
Thus a radial impact is most effective at keeping the van Vleck
determinant large. Grazing impacts lead to small values for the
van Vleck determinant.

Note that as the base point of the geodesic approaches the mouth
of the wormhole $\ell_+ \ell_- \to 0$, so that (again) $\Delta_\gamma(x,x)
\to 1$. The case $\theta = \pi/2$, which corresponds to the geodesic
having a tangential impact on the mouth of the wormhole, naively
leads to the indeterminate form $0/0$. To regularize this, recognize
that tangential impact $\theta = \pi/2$, occurs only if one very
carefully orients the two mouths of the wormhole to be facing away
from each other.  But it is impossible to hold the orientation of
the wormhole mouths completely fixed.  If nothing else, the Heisenberg
uncertainty principle provides a fundamental limitation $\Delta\theta
\Delta L \approx  \hbar$ relating the spread in orientation to the
spread in angular momentum of the wormhole mouth.  If one centers
the orientation on $\theta = \pi/2$, one has $\langle \sec^2\theta
\rangle \approx {\langle \theta^2 \rangle}^{-1} \approx (\Delta L
/\hbar)^{2} \equiv \langle J^2 \rangle$. Thus a ``regulated'' van
Vleck determinant, modified by orientational smearing may be taken
to be
\begin{equation}
\Delta_\gamma(x,x;\theta = \pi/2) \approx
 \left( 1 + {\ell_+\ell_-\over \ell} {2\over R}\right)^{-1}
  \left( 1 + {\ell_+\ell_-\over \ell}
             {2\langle J^2 \rangle \over R}\right)^{-1}_.
\end{equation}
This ``regulated'' quantity tends to $1$ as $x$ approaches the
throat of the wormhole, which solves the problem of what to do with
the indeterminate form.

%%%%%%%%%%%%%%%%%%%%%%%%%%%%%%%%%%%%%%%%%%%%%%%%%%%%%%%%%%%%%%%%%%%%%%%%%%%%%%
\subsection{Double pass: exact result}
%%%%%%%%%%%%%%%%%%%%%%%%%%%%%%%%%%%%%%%%%%%%%%%%%%%%%%%%%%%%%%%%%%%%%%%%%%%%%%

Consider a geodesic that starts at the point $x$, wraps through
the wormhole twice, and returns to the point $x$. In view of the
complexity of structure to be outlined below it is not particularly
enlightening to even contemplate non radially impacting geodesics.
Henceforth one specializes even further by assuming collinear motion
for the wormhole mouths ({\sl i.e.} radial approach or recession)
and restricting attention to ``central geodesics''. The geodesic
has three tangent vectors, $t_\pm$, and $t_0$. One each of the two
$t_\pm$ points towards each of the two wormhole mouths. $t_0$ points
from one wormhole mouth to another. Each of the two wormhole mouths
is characterized by a four-velocity $V_\pm$. There are two normals
to keep track of: $n_\pm$. Here $n_+$ is the normal to the wormhole
mouth with velocity $V_+$, evaluated at the point where the geodesic
from $x$ impacts that mouth. Similarly for $n_-$. One can define
quantities $\gamma_\pm = (1-\beta_\pm^2)^{-1/2}$ by
\begin{eqnarray}
t_\pm^\mu &&= \gamma_\pm\left(
\beta_\pm \; V_\pm^\mu + n_\pm^\mu
\right).  \\
t_0^\mu &&= \pm \gamma_\pm\left(
\beta_\pm \; V_\mp^\mu + n_\mp^\mu
\right).
\end{eqnarray}
Here, as was the case for the $N=1$ case, one has utilized the fact
that the tangent vectors are (for current purposes) spacelike unit
vectors. Because the geodesic must, by construction, pass through
the wormhole mouth in a smooth manner one must have the same
coefficients $\gamma_\pm$, and $\beta_\pm$, occurring in these
equations.  In particular, $t_\pm \cdot n_\pm = \pm t_0 \cdot{
n}_\mp = \gamma_\pm$. Because $V_\pm$ and $n_\pm$ are all coplanar,
\begin{equation}
g^\perp_{\mu\nu} =
(g_{\mu\nu} + V^\pm_\mu V^\pm_\nu - n^\pm_\mu n^\pm_\nu ).
\end{equation}
This obviates the otherwise messy technical requirement of keeping
track of precisely which of the $g^\perp$ one is dealing with, and
permits the radical simplification
\begin{equation}
[q^\pm_{\mu\nu}] =
\gamma_\pm \;{2\over R} \; g^\perp_{\mu\nu}.
\end{equation}
So finally
\begin{equation}
\Delta_\gamma(x,x)
= \left( 1
+  {s_+(s_- + s_0)\over s} {2\gamma_+\over R}
+  {s_-(s_+ + s_0)\over s} {2\gamma_-\over R}
+  {s_+ s_- s_0\over s}{4\gamma_+\gamma_-\over R^2} \right)^{-2}_.
\end{equation}
This is the promised exact result for an arbitrary central
geodesic that wraps twice through the wormhole.

One may still define the quantities $\ell_\pm \equiv \gamma_\pm
s_\pm = s_\pm \; (t_\pm \cdot n_\pm)$, and interpret these quantities
as the physical distance between the point $x$ and the relevant
wormhole mouth, as measured in the rest frame of that wormhole
mouth. The physical import of this result for the van Vleck
determinant is unfortunately nowhere near as transparent as that
derived for the case of a single pass.

%%%%%%%%%%%%%%%%%%%%%%%%%%%%%%%%%%%%%%%%%%%%%%%%%%%%%%%%%%%%%%%%%%%%%%%%%%%%%%
\subsection{Multiple passes: exact result}
%%%%%%%%%%%%%%%%%%%%%%%%%%%%%%%%%%%%%%%%%%%%%%%%%%%%%%%%%%%%%%%%%%%%%%%%%%%%%%

Most of the preparatory work for this section has already been
done. For a completely general geodesic the algebraic complexity
would be absolutely prohibitive.  Accordingly, it is useful to
specialize to the case of a completely smooth closed geodesic. One
that wraps around the wormhole $N$ times.

Invoking the preceding multi--pass calculation, and now using the
extended thin shell formalism to probe the Riemann tensor at the
wormhole throat, one has
\begin{equation}
s_0 [q^\mu{}_\nu]
= s_0 \gamma [\kappa^\mu{}_\nu]
= \ell_0 [\kappa^\mu{}_\nu]
= \ell_0
\left[ {\rm diag}\left\{0,(2/R_1),(2/R_2),0\right\}^\mu{}_\nu\right].
\end{equation}
The net result of the various factors of $\gamma$ has been to
quietly disappear, after conveniently converting the arc--length
along the geodesic into the physical distance between the wormhole
mouths, as measured in the rest frame of the wormhole mouths.

Substitution into the previous results show
\begin{eqnarray}
\Delta_\gamma^{-1}
&=& \left\{ 1 +  {2 p_+ p_- \ell_0\over R_1}  \right\} \times
    \left\{ 1 +  {2 p_+ p_- \ell_0\over R_2}  \right\}
\nonumber\\
&&\qquad \times
          \left\{ \sum_{n=0}^{N-1} {N+n\choose 2n+1}
          \left[{2\ell_0\over R_1}\right]^n \right\} \times
	 \left\{ \sum_{n=0}^{N-1} {N+n\choose 2n+1}
          \left[{2\ell_0\over R_2}\right]^n \right\} \times N^{-2}_.
\end{eqnarray}
Equivalently
\begin{eqnarray}
\Delta_\gamma
&=& \left\{ 1 +  {2 \ell_+ \ell_- \over \ell_0 R_1}  \right\}^{-1}
    \times
    \left\{ 1 +  {2 \ell_+ \ell_- \over \ell_0 R_2}  \right\}^{-1}
\nonumber\\
&&\qquad \times N^2 \times
          \left\{ \sum_{n=0}^{N-1} {N+n\choose 2n+1}
          \left[{2\ell_0\over R_1}\right]^n \right\}^{-1} \times
	  \left\{ \sum_{n=0}^{N-1} {N+n\choose 2n+1}
          \left[{2\ell_0\over R_2}\right]^n \right\}^{-1}_.
\end{eqnarray}
Note that if $\ell_0 \gg R_{(1,2)}$
\begin{eqnarray}
\Delta_\gamma
&\approx&
    \left\{ 1 +  {2 \ell_+ \ell_- \over \ell_0 R_1}  \right\}^{-1}
    \times
    \left\{ 1 +  {2 \ell_+ \ell_- \over \ell_0 R_2}  \right\}^{-1}
\nonumber\\
&&\qquad \times N^2 \times
         \left\{ { 4 \ell_0^2\over R_1 R_2} \right\}^{N-1}_.
\end{eqnarray}
Again, this result should be compared to eqs. (64) and (48) of Kim
and Thorne~\cite{Kim-Thorne}.  The present result is compatible
with, and at least along the central geodesic, a extension of the
Kim-Thorne results.

%%%%%%%%%%%%%%%%%%%%%%%%%%%%%%%%%%%%%%%%%%%%%%%%%%%%%%%%%%%%%%%%%%%%%%%
\subsection{Discussion: isolated wormholes}
%%%%%%%%%%%%%%%%%%%%%%%%%%%%%%%%%%%%%%%%%%%%%%%%%%%%%%%%%%%%%%%%%%%%%%%

Having all these computations of the van Vleck determinant in
isolated wormhole spacetimes in hand, I turn to the issue of
Hawking's chronology protection conjecture. Recall that vacuum
polarization effects disrupt the internal structure of the wormhole
once {\em any} of the closed geodesics through the wormhole is shorter
than
\begin{equation}
s\vert_{\rm disrupt} \approx
\max_{\gamma}  \left\{
        \max_x \left\{
	         \Delta_{\gamma}^{1/8}(x,x) \sqrt{\pi \ell_P R}
	       \right\}
	       \right\}_.
\end{equation}
The maximization runs over all closed geodesics $\gamma$ with base
point $x$ on the throat of the wormhole.

In particular, let $\gamma_1$ be a geodesic that wraps only once
through an isolated wormhole, then all the preceding calculations
agree that when $x$ lies on the throat of the wormhole
\begin{equation}
\Delta_{\gamma_1}(x,x) = 1.
\end{equation}
As the base point $x$ moves away from the throat of the wormhole,
the van Vleck determinant may, and often will, rapidly fall off to
zero. Nevertheless, for wormhole disruption effects, it is the
behaviour at the throat that is relevant. One now has
\begin{equation}
s\vert_{\rm disrupt}
\approx  \sqrt{\pi \ell_P R} \gg \ell_P.
\end{equation}
Which is what I wanted to prove.

%%%%%%%%%%%%%%%%%%%%%%%%%%%%%%%%%%%%%%%%%%%%%%%%%%%%%%%%%%%%%%%%%%%%%%%%%%%%%%
\section{THE ROMAN CONFIGURATION}
%%%%%%%%%%%%%%%%%%%%%%%%%%%%%%%%%%%%%%%%%%%%%%%%%%%%%%%%%%%%%%%%%%%%%%%%%%%%%%

The analysis of multiple wormhole configurations is a topic of
considerable additional complexity. It is relatively easy to think
up a putative time machine built out of two or more wormholes, each
of which taken in isolation is not itself a time machine. The
simplest such configuration was considered by Tom
Roman~\cite{GR13,Roman,Thorne}, and uses two wormholes ({\em
vide}~\cite{Visser-baby}).

For example, for the time being ignore the internal structure of
the two wormholes and simply model them by identifying two
world--lines. Let wormhole \#1 be defined by
\begin{equation}
(t,0,0,0) \equiv (t+T_1,0,0,\ell).
\end{equation}
As long as the distance between the wormhole mouths $\ell$ is
greater than the time shift $T_1$, wormhole \#1 does not, in and
of itself, constitute a time machine. Now add a second wormhole.
Let wormhole \#2 be offset a distance $\zeta$ along the $x$--axis.
That is
\begin{equation}
(t,\zeta,0,\ell) \equiv (t+T_2,\zeta,0,0).
\end{equation}
Again, as long as $\ell > T_2$, wormhole \#2 does not, in and of
itself, describe a time machine. Now consider the null trajectory
\begin{eqnarray}
&&(0,0,0,0) \equiv
(T_1,0,0,\ell)
\nonumber\\
&& \qquad \to (T_1+\zeta,\zeta,0,\ell) \equiv
(T_1+\zeta+T_2,\zeta,0,0)
\nonumber\\
&& \qquad \qquad \to
(T_1+\zeta+T_2+\zeta,0,0,0).
\end{eqnarray}
This future--pointing null trajectory has returned to its spatial
starting point in a total time $\Delta T = T_1+T_2+2\zeta$. This
total time shift can easily be arranged to be negative. (Example:
$\ell = 3\zeta$, $T_1 = T_2 = -2\zeta$, $\Delta T = -2\zeta$.) We
note that $\Delta T > -2\ell + 2\zeta$, so that a necessary condition
for this compound ``Roman configuration'' to form a time machine
is $\ell > \zeta$. (The individual wormholes should permit jumps
across the universe that are longer than the ``offset'' distance
between wormhole \#1 and wormhole \#2.) The maximum possible size
of the backward time jump is $2\ell - 2\zeta$. One can add some
internal structure to the wormhole by making the mouths spherically
symmetric of radius $R$, as long as $R \ll \zeta$ this will not
disturb the previous discussion.

%%%%%%%%%%%%%%%%%%%%%%%%%%%%%%%%%%%%%%%%%%%%%%%%%%%%%%%%%%%%%%%%%%%%%%%%%%%%%%
\subsection{van Vleck determinant: Roman configuration}
%%%%%%%%%%%%%%%%%%%%%%%%%%%%%%%%%%%%%%%%%%%%%%%%%%%%%%%%%%%%%%%%%%%%%%%%%%%%%%

Evaluation of the van Vleck determinant for ``Roman configuration''
spacetimes is subtle.  One shall soon see that it seems that vacuum
polarization effects in such multiple wormhole putative time machines
can, with a suitably bizarre choice of parameters,  be arranged to
be arbitrarily small at the onset of Planck scale physics.  This
is disturbing. (This property of Roman configuration wormhole
systems has also been noted by Lyutikov~\cite{Lyut}, see the
announcement in~\cite{GR13}.)

Start the analysis by placing the wormhole mouths in general positions.
All mouths are taken to be at rest with respect to one another,
and wormhole \#1 is described by the identification
\begin{equation}
(t,\vec x^{in}_1) \equiv (t+T_1,\vec x^{out}_1).
\end{equation}
Define $\ell_1 \equiv || \vec x^{out}_1 - \vec x^{in}_1||$, since
one does not wish this wormhole to be a time machine in its own right
$|T_1| < \ell_1$.  For wormhole \#2 one  simply copies all of these
definitions, for example
\begin{equation}
(t,\vec x^{in}_2) \equiv (t+T_2,\vec x^{out}_2).
\end{equation}
Now consider the following closed curve $C(t_0)$
\begin{equation}
(0,\vec x^{in}_1) \equiv (T_1,\vec x^{out}_1) \to
(t_0,\vec x^{in}_2) \equiv (t_0+T_2,\vec x^{out}_2) \to
(0,\vec x^{in}_1).
\end{equation}
Here $t_0$ is a parameter that is for the time being arbitrary. We
shall adjust $t_0$ in such a manner as to make $C$ a geodesic. The
arc length along the curve $C(t_0)$ is
\begin{equation}
s[C] = \sqrt{ (\vec x^{out}_1 - \vec x^{in}_2)^2 - (T_1-t_0)^2 }
     + \sqrt{ (\vec x^{out}_2 - \vec x^{in}_1)^2 - (t_0+T_2)^2 }.
\end{equation}
To simplify life, introduce the notation
\begin{eqnarray}
\ell_{1\to2} &\equiv& || \vec x^{out}_1 - \vec x^{in}_2||, \\
\ell_{2\to1} &\equiv& || \vec x^{out}_2 - \vec x^{in}_1||.
\end{eqnarray}
The arc length of the curve $C$ is extremized when
\begin{equation}
{\partial s[C(t_0)] \over \partial t_0} \equiv
{ T_1-t_0\over \sqrt{ (\ell_{1\to2})^2 - (T_1-t_0)^2} } -
{ T_2+t_0\over \sqrt{ (\ell_{2\to1})^2 - (T_2+t_0)^2} } = 0.
\end{equation}
This equation has the parametric solution
\begin{eqnarray}
T_1 - t_0 &=& k \ell_{1\to2}, \\
T_2 + t_0 &=& k \ell_{2\to1}, \\
T_1 + T_2 &=& k ( \ell_{1\to2} + \ell_{2\to1} ).
\end{eqnarray}
So the arc--length of the closed geodesic of current interest, one
that wraps once through the two wormhole system, obtained by setting
$t_0$ to its critical value, is
\begin{equation}
s[\gamma] = \sqrt{ ( \ell_{1\to2} + \ell_{2\to1} )^2 - (T_1+T_2)^2 }
\end{equation}
A putative time machine forms once $|T_1+T_2| >  \ell_{1\to2} +
\ell_{2\to1}$. This is a necessary and sufficient condition. But
because one does not wish the individual wormholes to be time machines
$|T_1|<\ell_1$, $|T_2| < \ell_2$. Consequently $|T_1+T_2| < |T_1|
+ |T_2| < \ell_1 + \ell_2$. So another necessary (but not sufficient)
condition on the formation of a ``two wormhole time machine'' is
$ \ell_{1\to2} + \ell_{2\to1} < \ell_1 + \ell_2$.  This means that
the net distance ``jumped'' through the two wormholes has to exceed
the total distance travelled in going from one wormhole to the other.

One may also consider the behaviour of a future pointing null curve
(not a closed curve) that threads the two wormholes
\begin{eqnarray}
&&
(0,\vec x^{in}_1) \equiv
(T_1,\vec x^{out}_1)
\nonumber\\
&& \qquad \to
(T_1+\ell_{1\to2},\vec x^{in}_2) \equiv
(T_1+\ell_{1\to2}+T_2,\vec x^{out}_2)
\nonumber\\
&& \qquad \qquad \to
(T_1+\ell_{1\to2}+T_2+\ell_{2\to1},\vec x^{in}_1).
\end{eqnarray}
This curve returns to its spatial starting point in total time
$\Delta T \equiv T_1+T_2+\ell_{1\to2}+\ell_{2\to1}$. As per the
preceding discussion this can easily be negative. The virtue of
this type of analysis is that it is now clear that the maximum
possible size of the (single trip) backward time jump is $\Delta
T_{max} \equiv \ell_1 + \ell_2 - \ell_{1\to2} - \ell_{2\to1}$.

Return to the general arguments given in the early portions of this
paper. For a geodesic that makes a single pass through the mouths
of two distinct wormholes ($N=2$), one may still write down the
closed form expression
\begin{eqnarray}
A^\mu{}_\nu(s)
&=&  s \delta^\mu{}_\nu  \nonumber\\
&& + \Theta(s-s_1)
   \left\{(s-s_1) [q(1)^\mu{}_\nu] s_1 \right\}
\nonumber\\
&& + \Theta(s-s_2)
    \left\{(s-s_2) [q(2)^\mu{}_\nu] s_2 \right\}
\nonumber\\
&&+ \Theta(s-s_2)
    \left\{(s-s_2) [q(2)^\mu{}_\rho] (s_2-s_1)
                             [q(1)^\rho{}_\nu] s_1 \right\}_.
\end{eqnarray}
Particularize to a closed geodesic, so that $x=y$. Let the base
point $x$ lie  on the mouth of wormhole \#1. Then $s_1=0$, and we
may define $s_{1\to2} = s_2$, $s_{2\to1} = s - s_2$, so that $s =
s_{1\to2} + s_{2\to1}$. Thus
\begin{equation}
\Delta_\gamma(x,x)^{-1}
= \det\left(
\delta^\mu{}_\nu
+  {s_{1\to2} \; s_{2\to1}\over s} [q(2)^\mu{}_\nu] \right).
\end{equation}

Now invoke the extended thin shell formalism to estimate
$[q(2)^\mu{}_\nu]$. For radial impact the previous arguments give
$[q(2)^\mu{}_\nu] = (t\cdot n) [\kappa(2)^\mu{}_\nu] = \gamma
(2/R_2) [g^\perp(2)^\mu{}_\nu]$. Here $\kappa(2)$ denotes the
discontinuity in the second fundamental form at the throat of
wormhole \#2. Assuming spherical symmetry I have taken $R_2$ to
denote the radius of the throat. The factor $\gamma = (t\cdot n)$
is easily seen to be
\begin{equation}
\gamma = {\ell_{1\to2}\over s_{1\to2}}
       = {\ell_{2\to1}\over s_{2\to1}}
       = {\ell_{1\to2} + \ell_{2\to1}\over s_{1\to2} + s_{2\to1} }.
\end{equation}
The net result is now
\begin{equation}
\Delta_\gamma(x,x) =
\left\{ 1 +
  {2 \ell_{1\to2} \; \ell_{2\to1} \over
   ( \ell_{1\to2} + \ell_{2\to1} ) R_2}
\right\}^{-2}.
\end{equation}
This is compatible with the discussion on page 311 of~\cite{GR13},
and for the case of a single trip through the compound system,
generalizes that discussion to arbitrary placement of the wormhole
mouths. See also~\cite{Lyut}.

%%%%%%%%%%%%%%%%%%%%%%%%%%%%%%%%%%%%%%%%%%%%%%%%%%%%%%%%%%%%%%%%%%%%%%%
\subsection{Discussion: multiple wormholes}
%%%%%%%%%%%%%%%%%%%%%%%%%%%%%%%%%%%%%%%%%%%%%%%%%%%%%%%%%%%%%%%%%%%%%%%

For a ``Roman configuration'' the wormhole disruption criterion
may be given as
\begin{eqnarray}
s\vert_{\rm disrupt}
&\approx&
\left[ R_2
       \left( {1\over\ell_{1\to2}} +  {1\over\ell_{2\to1}} \right)
\right]^{1/4}
\sqrt{\pi \ell_P R_1} \\
&\approx&
\ell_P \left[ {R_1^2 R_2 \over \ell_P^2}
              \left({1\over\ell_{1\to2}}+{1\over\ell_{2\to1}} \right)
       \right]^{1/4}
\end{eqnarray}
The best possible case for the possibility of time travel is obtained
if one makes $\Delta_\gamma(x,x)$ and hence $s\vert_{\rm disrupt}$
as small as possible, and pushes the disruption scale down into
the Planck slop. This may be achieved by making $R$ as small as
possible, and $\ell$ as large as possible. If one tries to force
$s\vert_{\rm disrupt} < \ell_P$ one acquires a constraint on the
relevant wormhole parameters.  Indeed
\begin{equation}
R_1^2 R_2
<
\ell_P^2 \;
  {\ell_{1\to2} \ell_{2\to1} \over \ell_{1\to2} + \ell_{2\to1} }
<
\ell_P^2 \; \max(\ell_{1\to2}, \ell_{2\to1})
<
\ell_P^2 R_{\rm universe}.
\end{equation}
Here the best possible case for time travel has been made by relaxing
the separation of the wormholes as much as possible --- surely the
radius of the universe is a good upper bound on the distance between
the wormholes. Take $R_{\rm universe} \approx 3 \hbox{\rm\ Giga--parsecs}
\approx 6 \times 10^{60} \ell_P$. Then
\begin{equation}
R_1^2 R_2 <
6 \times (10^{20} \ell_P)^3
\approx (3 \times 10^{-15} m)^3.
\end{equation}
So even with these ludicrously large separations between the wormhole
mouths, one can only push the disruption scale down into the Planck
slop by building the putative time machine out of even ludicrously
smaller ``traversable'' wormholes with a radius of the order of
{\em femtometres}. Since any would be time traveller would have to
traverse the distance between wormhole \#1 and wormhole \#2 in
normal space, he/she/it had better also be patient --- a lifetime
on the order of Giga--years would be appropriate.

If one attempts to build a ``two wormhole time machine'' on a more
modest human scale one might take as a good bound $ \max(\ell_{1\to2},
\ell_{2\to1}) < 1 {\rm\ AU} \approx 9 \times 10^{45} \ell_P$. In this case
\begin{equation}
R_1^2 R_2 <
(2 \times 10^{15} \ell_P)^3
\approx (10^{-20} m)^3
\approx \left( {\hbar\over 20 {\rm TeV/c} } \right)^3.
\end{equation}
So even if one lays hands on a couple of wormholes, and initiates
suitable solar system scale engineering projects, any decent sized
wormhole will be disrupted by vacuum polarization effects before
time travel is achieved. If the wormholes in question are small
enough one can push the disruption scale down below the Planck
regime. This, of course does not mean one has proved that time
travel is actually possible --- all it means is that one has
``fine--tuned'' the system sufficiently to be in a parameter regime
where one cannot trust even rudimentary calculations. Even if one
were to then succeed in building a closed timelike loop, the small
size of the wormhole would preclude anything short of a $20$ TeV
quantum from getting through. This implies that one would need two
SSC scale accelerators just to get a one--bit message through the
putative time machine. Even if all of these constraints are satisfied,
any single trip is limited to a maximum time jump of less than $(
1 {\rm\ AU}/c ) \approx 8 {\rm\ minutes}$. This does not seem to
be a useful workable recipe for studying tomorrow's Wall Street
Journal.

%%%%%%%%%%%%%%%%%%%%%%%%%%%%%%%%%%%%%%%%%%%%%%%%%%%%%%%%%%%%%%%%%%%%%%%
\section{CONCLUSIONS}
%%%%%%%%%%%%%%%%%%%%%%%%%%%%%%%%%%%%%%%%%%%%%%%%%%%%%%%%%%%%%%%%%%%%%%%

The recent flurry of interest in traversable wormholes, with
concomitant realization that it {\em appears} to be easy to turn
a traversable wormhole into a time machine, has led to a great deal
of interest.  The issue of time travel is one that should be faced
squarely as it cuts at the very foundations of what is currently
believed about the structure of physics and the nature of the universe
(multiverse?).  While many authors are busy making their peace with
the notion of time travel and its associated paradoxes, Hawking
has promulgated his {\em chronology protection conjecture} whereby
time travel is forbidden and the universe is made ``safe for
historians''.

In an earlier paper~\cite{Visser-CPC} I argued that the universe
appears to defend causality through a strategy of ``defense in
depth''. This paper addresses the penultimate line of defense,
vacuum polarization effects. This line of defense can only be
formulated if one is in the semiclassical regime --- a well defined
background spacetime manifold is required, and one proceeds to
perform quantum field theoretic calculations on that background.
Vacuum polarization effects become infinite as one gets close to
forming a time machine, thereby disrupting the traversable wormhole(s)
used in the attempt to construct the time machine, and so aborting
the formation of the time machine.

The technical aspects of this paper boil down to a minor agony of
calculations of the van Vleck determinant in suitable model spacetimes
that I believe mimic the essential features of traversable wormholes.

For putative time machines constructed out of a single wormhole,
the evidence is conclusive that wormhole disruption occurs long
before one enters the Planck regime.

For putative time machines constructed out of a two wormholes
arranged in a ``Roman configuration'', the situation is messier.
If the wormhole sizes and separations are human scale, then  wormhole
disruption occurs long before one enters the Planck regime.  For
suitably obtuse (not traversable by humans) choices of wormhole
size and location one can push the disruption scale down below the
Planck slop. Even begging the question of whether or not one can
build or acquire traversable wormholes in the first place, this
does {\em not} mean that a ``Roman configuration'' of wormholes
can be used to build a time machine.  All it means is that one's
limited ability to calculate in the semiclassical regime has been
completely obviated by entry into the uncharted wasteland of quantum
gravity.

Perhaps the best attitude to take towards Planck slop effects on
the {\em chronology protection conjecture} is that ultimately it
should be taken to be an axiom, rather than attempting to prove it
by calculation.

AXIOM: Quantum gravity, whatever it may be, is causal.

%%%%%%%%%%%%%%%%%%%%%%%%%%%%%%%%%%%%%%%%%%%%%%%%%%%%%%%%%%%%%%%%%%%%%%%%
\acknowledgements
%%%%%%%%%%%%%%%%%%%%%%%%%%%%%%%%%%%%%%%%%%%%%%%%%%%%%%%%%%%%%%%%%%%%%%%%

I wish to thank Carl Bender, Kip Thorne, and Tom Roman for useful
discussions.  I also wish to thank Ian Redmount and Wai--Mo Suen
for their helpful comments and the Aspen Center for Physics for
its hospitality during an early phase of this work.  This research
was supported by the U.S. Department of Energy.

%%%%%%%%%%%%%%%%%%%%%%%%%%%%%%%%%%%%%%%%%%%%%%%%%%%%%%%%%%%%%%%%%%%%%%%%

%%%%%%%%%%%%%%%%%%%%%%%%%%%%%%%%%%%%%%%%%%%%%%%%%%%%%%%%%%%%%%%%%%%%%%%%
\end{document}